\documentclass[12pt,letterpaper]{article}
\usepackage{jheppub,amsmath,braket}
\usepackage{xcolor}
\usepackage{graphicx}
\DeclareMathOperator{\arccosh}{arccosh}
\DeclareMathOperator{\tr}{tr}

\title{Gravity/Ensemble Duality}
\author[a,b]{Raphael Bousso}
\author[a]{and Elizabeth Wildenhain}
\affiliation[a]{Center for Theoretical Physics and Department of Physics,\\
University of California, Berkeley, CA 94720, U.S.A.}
\affiliation[b]{Lawrence Berkeley National Laboratory, Berkeley, CA 94720, U.S.A.}
\emailAdd{bousso@berkeley.edu}
\emailAdd{elizabeth\_wildenhain@berkeley.edu}

\abstract{For the first time, a gravitational calculation was recently shown to yield the Page curve for the entropy of Hawking radiation, consistent with unitary evolution. However, the calculation takes as essential input Hawking's result that the radiation entropy becomes large at late times. We call this apparent contradiction the state paradox. We exhibit its manifestations in standard and doubly-holographic settings, with and without an external bath. We clarify which version(s) of the Ryu-Takayanagi prescription apply in each setting. We show that the two possible homology rules in the presence of a braneworld generate a bulk dual of the state paradox. The paradox is resolved if the gravitational path integral computes averaged quantities in a suitable ensemble of unitary theories, a possibility supported independently by several recent developments.}% The ensemble average of the entropy need not agree with the entropy of the ensemble-averaged state of the radiation.}

\begin{document}
\maketitle

\section{Introduction}
\label{intro}

The black hole information paradox~\cite{Polchinski:2016hrw} is a conflict between quantum mechanics and general relativity. By the principle of unitarity, quantum information should be preserved in a scattering process that returns all energy to a distant observer. A pure in-state should evolve to a pure out-state. Hawking's calculation of black hole radiation~\cite{Haw74,Haw76}, however, implies that only the energy is returned, but not the information. For decades, the only concrete evidence against information loss came from an indirect argument: assuming the AdS/CFT correspondence~\cite{Mal97}, the S-matrix can be computed in the CFT and so must be unitary.

\paragraph{Unitarity from Ryu-Takayanagi} For the first time, a purely bulk calculation was recently found to support unitarity~\cite{Pen19,AEMM}. This challenges Hawking's conclusion directly, rather than through an asserted duality. The new analysis does not identify an error in Refs.~\cite{Haw74,Haw76}; in fact, it uses Hawking's calculation. But it asks a different question, which leads to a different conclusion.

\begin{figure}
    \centering
    \includegraphics[width=0.73\textwidth]{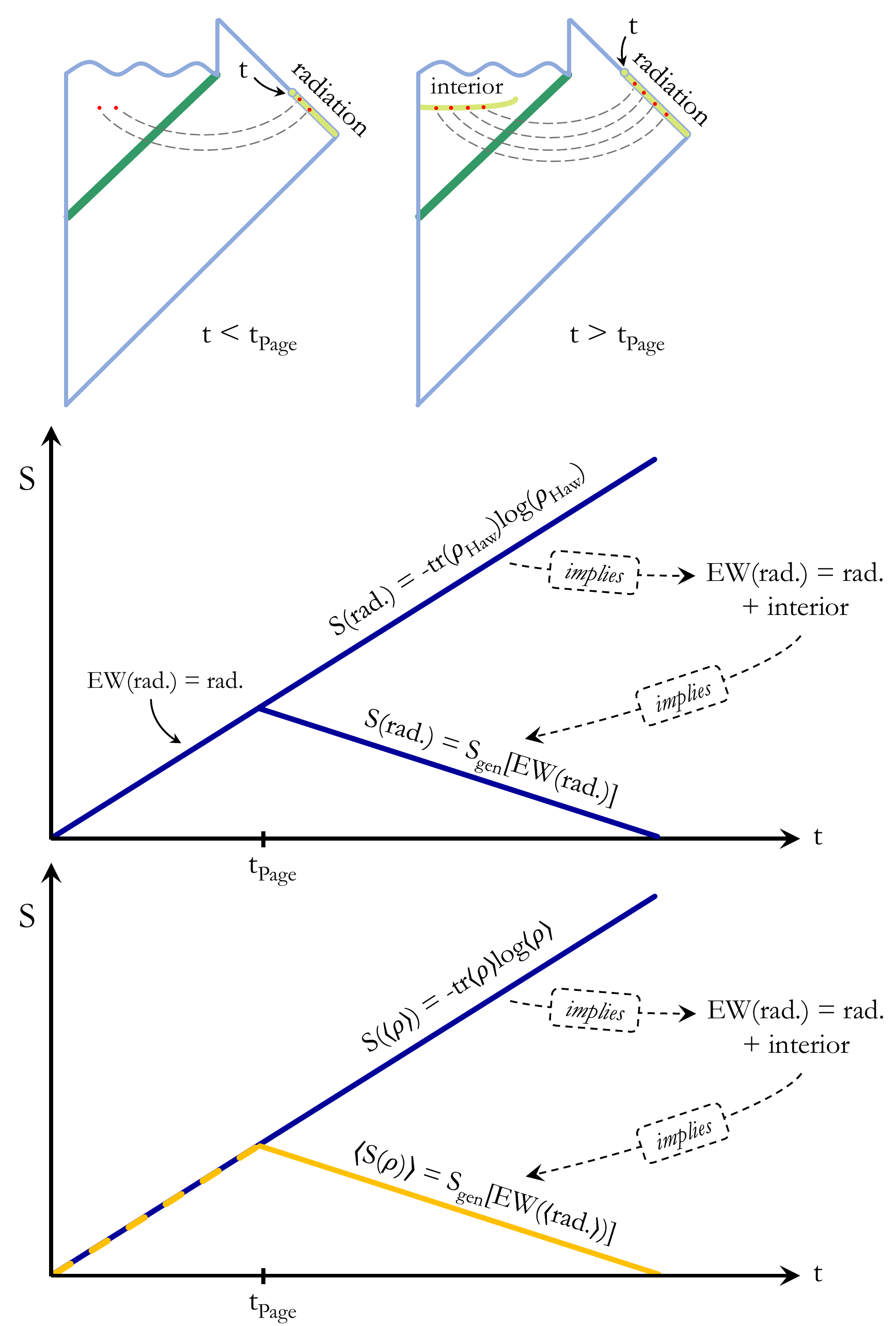}
    \caption{{\em Top:} Penrose diagrams for an evaporating black hole. The light green region is the entanglement wedge of the radiation that has arrived at infinity before (left) and after (right) the Page time. {\em Middle:} State paradox. The RT prescription yields the Page curve for the entropy of the radiation, but only if the same entropy is assumed to follow Hawking's rising curve when determining the entanglement wedge. {\em Bottom:} Resolution of the state paradox by gravity/ensemble duality. The ensemble-averaged state is mixed, and its entropy follows Hawking's curve. The ensemble-averaged entropy follows the Page curve.}
    \label{fig:intro}
\end{figure}

Hawking asked about the quantum state of the black hole radiation and found it to be a thermal state, $\rho_{\rm Haw}(t)$. Its von Neumann entropy, 
\begin{equation}
    S=-\tr \rho_{\rm Haw}\log\rho_{\rm Haw}~,
    \label{vn}
\end{equation} 
rises monotonically as more radiation is produced. When the black hole is fully evaporated, $S$ will be of order $\mathcal{A}_0/4G$, where $\mathcal{A}_0$ is the initial black hole area. See Fig.~\ref{fig:intro} (middle subfigure, upper graph).

By contrast, Refs.~\cite{Pen19,AEMM} ask only about the entropy of the radiation, not its state. The entropy $S$ is computed not via Eq.~\eqref{vn}, but as the analytic continuation of the $n$-th Renyi entropy to $n=1$. In the presence of gravity, this method is compactly encoded~\cite{LewMal13} in the Ryu-Takayanagi (RT) prescription~\cite{RyuTak06, RyuTak06b, HubRan07, FauLew13, EngWal13}. In Sec.~\ref{shb}, we will give a precise definition of the RT prescription in the setting of Refs.~\cite{Pen19,AEMM}. Schematically,
\begin{equation}
    S(\mbox{radiation}) = S_{\rm gen}[\mbox{EW(radiation)}]~,
    \label{rtsimple}
\end{equation}
where EW denotes a region called the entanglement wedge, and
\begin{equation}
    S_{\rm gen}(\mbox{EW}) =\frac{\mathcal{A}(\partial \mbox{EW})}{4G}+ S(\mbox{EW})
    \label{sgen}
\end{equation}
is the generalized entropy. Here $\partial$ denotes the boundary of a region, $\mathcal{A}$ denotes the area, and $G$ is Newton's constant. 

The spacetime and its matter fields are computed using Hawking's approach, semiclassical gravity. But by using Eq.~\eqref{rtsimple} instead of Eq.~\eqref{vn}, one finds that $S(\mbox{radiation})$ follows the ``Page curve'' demanded by unitary evolution. It rises until the Page time, $t_{\rm Page}$, when the black hole and radiation entropies are equal. Then $S(\mbox{radiation})$ falls, ultimately vanishing when the evaporation is complete. See Fig.~\ref{fig:intro} (middle subfigure, lower graph).

Before the Page time, EW(radiation) is the radiation itself (Fig.~\ref{fig:intro}, left Penrose diagram). The RT prescription adds nothing new; the entropy rises because it does so in Hawking's calculation. After the Page time (right Penrose diagram), a minimality condition in the definition of the entanglement wedge implies\footnote{When the minimality condition results in an island, it has been called the ``island rule''~\cite{AMMZ}. However, this is not a new rule nor a modification of RT. The existence of islands after the Page time already follows from the RT prescription in the final form given to it by Engelhardt and Wall~\cite{EngWal13}.} that EW(radiation) contains both the radiation and a disconnected ``island,'' the black hole interior:
\begin{equation}
    \mbox{EW(radiation)} = \mbox{radiation} + \mbox{black hole interior}~~~~~~(t>t_{\rm Page})~.
\end{equation}
In Hawking's analysis, the interior together with the Hawking radiation are in a pure state. Hence the von Neumann entropy $S(\mbox{EW})$ vanishes, and only the area of the boundary of the island contributes. This boundary is approximately the black hole horizon, so
\begin{equation}
    S_{\rm gen}[\mbox{EW(radiation)}] = \frac{\mathcal{A}(\rm horizon)}{4G}~~~~~~(t>t_{\rm Page})~.
    \label{sgr}
\end{equation}
The horizon area decreases as the black hole evaporates, yielding the falling part of the Page curve.

\paragraph{State Paradox} The breakthrough of Refs.~\cite{Pen19,AEMM} involves an apparent paradox: it makes use of Hawking's result that $S(\mbox{radiation})$ increases monotonically for all times, in order to reach the final conclusion that it does not. Through Eq.~\eqref{sgr}, the radiation appears on both sides of Eq.~\eqref{rtsimple}. Hawking's Eq.~\eqref{vn} is invoked in evaluating its entropy on the RHS of Eq.~\eqref{rtsimple}. Thus on the RHS, $S(\mbox{radiation})$ (without the island) follows Hawking's monotonically increasing curve. This is a crucial ingredient, because it triggers the inclusion of the black hole interior in EW(radiation) after the Page time. On the LHS, $S(\mbox{radiation})$ then follows the Page curve. 

This is a contradiction. The S-matrix is an observable, so the state of the Hawking radiation cannot be ambiguous. Therefore, its von Neumann entropy cannot have two different values.%
\footnote{One might be tempted to declare that S computed from Eq.~\eqref{vn} is only a coarse-grained entropy (even though no coarse-graining is manifest in Hawking’s calculation). But the second term on the right side of Eq.~\eqref{sgen} is a fine-grained von Neumann entropy, and it is this fine-grained entropy that determines EW(radiation) in Eq.~\eqref{rtsimple}. The island that leads to the Page curve can only be included if the \textit{fine-grained} entropy of the radiation continues to grow after the Page time. This is achieved by taking Hawking’s calculation seriously at this step in the calculation, as a fine-grained entropy. Moreover, if the Page curve was assumed from the beginning, then the smooth horizon shown in the top diagrams in Fig.~\ref{fig:intro} would be inconsistent~\cite{AMPS} and so cannot enter the analysis at all. Finally, rejecting Eq.~\eqref{vn} as a fine-grained entropy would amount to putting in the Page curve by hand. With the Page curve for the radiation as input, Eq.~\eqref{rtsimple} would reproduce the Page curve trivially as an identity, not by inclusion of an island.}
We will call this contradiction the {\em state paradox}.

One possible resolution of the paradox is that the RT prescription is an uncontrolled approximation. Our confidence in the RT prescription derives from its success in the context of AdS/CFT, where the CFT entropy can often be independently computed and shown to agree. However, under certain assumptions, the RT prescription follows directly from a bulk path integral computation~\cite{LewMal13}, evaluated in the saddle point approximation. It is obtained as the analytic continuation to $n=1$ of the $n$-th Renyi entropies of the radiation, which can be computed from a path integral using the replica trick. After the Page time, one finds that the dominant saddle point has wormholes connecting the replicas~\cite{Penington:2019kki,Almheiri:2019qdq}; see Ref.~\cite{Almheiri:2020cfm} for a pedagogical review.

Thus the RT prescription has nothing to do with AdS/CFT; the nonperturbative completeness of the CFT is not used. RT can be applied even in asymptotically flat space, for example to compute the entropy of radiation that has arrived at the conformal boundary~\cite{Hartman:2020swn,Gautason:2020tmk}. RT is an advanced analogue of the Euclidean computation of the thermodynamic entropy of a black hole by Gibbons and Hawking~\cite{GibHaw77b}. It cleverly extracts information about the full quantum gravity theory from a path integral approximation. 

This is not a controlled approximation. It need not agree with the full quantum gravity theory, and when it does, it need not be self-consistent. This could explain the state paradox: perhaps Eq.~\eqref{rtsimple} just happens to compute the correct statistical entropy from the incorrect state. (See Ref.~\cite{Akers:2019nfi} for a discussion of related ideas.) And one day, perhaps, an even more sophisticated application of the Euclidean gravity path integral will be shown to yield the correct state of the Hawking radiation.

\paragraph{Gravity/Ensemble Duality} A different, intriguing possibility is that there exists a novel kind of duality: between an appropriately defined version of the gravitational path integral, and an ensemble of quantum mechanical theories without gravity. This can resolve the state paradox~\cite{Bousso:2019ykv}. According to this proposal, $S(\mbox{radiation})$ takes two different values on the two sides of Eq.~\eqref{rtsimple} because it is not the same quantity on the two sides. 

On the left side, it is the ensemble average of the entropy, so we should replace $S(\mbox{radiation})\to \braket{S(\rho)}$. See Fig.~\ref{fig:intro} (bottom subfigure, lower graph). On the right side, the entanglement wedge is determined from the entropy of the ensemble-averaged state of the radiation, $S(\braket{\rho})$. See Fig.~\ref{fig:intro} (bottom subfigure, upper graph). Because the von Neumann entropy is not a linear function of the state, generically $\braket{S(\rho)}\neq S(\braket{\rho})$. 

We now describe this proposal in more detail. Let $\nu$ label unitary theories, each capable of computing a pure Hawking radiation out-state from any pure in-state. Let 
\begin{equation}
    \braket{x}\equiv \int d\nu\, c(\nu)\, x^{(\nu)}
    \label{average}
\end{equation} 
denote an appropriately weighted average of the quantity $x$ computed in the different theories. Let $\rho_{\rm in}$ be the initial state before the black hole forms, and $\rho_{\rm out}^{(\nu)}$ be the final state of the radiation when the black hole has fully evaporated. Since each theory is unitary, we have 
\begin{equation}
S(\rho_{\rm out}^{(\nu)})=0 ~~~\mbox{for all}~ \nu~,
\label{uninu}
\end{equation} 
and hence
\begin{equation}
    \braket{S(\rho_{\rm out})}=0~.
\end{equation}
But in general, the final states $\rho_{\rm out}^{(\nu)}$ will be different in different theories. We now assume that their ensemble average is the thermal state predicted by Hawking:
\begin{equation}
    \braket{\rho_{\rm out}}=\rho_{\rm Haw}~.
    \label{hawkout}
\end{equation}
With these assumptions, Hawking's calculation computes the averaged out-state $\braket{\rho_{\rm out}}$; and in the same spacetime, the RT prescription correctly computes the averaged entropy:
\begin{equation}
    \braket{S[\rho_{\rm out}]} = S_{\rm gen}[\mbox{EW}(\braket{\rho_{\rm out}})]~.
    \label{rtensembleend}
\end{equation}
Moreover, this holds at all times. Let $\rho(t)=\tr_{>t} \rho_{\rm out}$ be the state of the radiation subsystem that has escaped to a distant region by the time $t$. The ensemble version of the RT prescription, Eq.~\eqref{rtsimple} states that
\begin{equation}
    \braket{S[\rho(t)]} = S_{\rm gen}[\mbox{EW}(\braket{\rho(t)})]~.
    \label{rtensemble}
\end{equation}
No contradiction arises. The ensemble average of the entropy will follow the Page curve, while the entropy of the ensemble average follows Hawking's curve.

The state paradox and its resolution by gravity/ensemble duality was first described in a slightly different setting~\cite{Bousso:2019ykv}, which we will review in Sec.~\ref{sh}. Another compelling argument for gravity/ensemble duality comes from the fact that the partition function on multiple copies of a boundary need not factorize when it is computed from a bulk gravity dual, because connected geometries can contribute~\cite{Penington:2019kki,Almheiri:2019qdq}. It would be interesting to understand the detailed relation between these arguments. 

The duality between JT gravity~\cite{Teitelboim:1983ux, Jackiw:1984je} and a random matrix ensemble furnishes an important concrete example of gravity/ensemble duality~\cite{Saad:2018bqo,Saad:2019lba,Stanford:2019vob,Saad:2019pqd,Witten:2020wvy}. Recently, an average over certain two-dimensional CFTs was shown to exhibit properties of an exotic three-dimensional gravity theory~\cite{Afkhami-Jeddi:2020ezh,Maloney:2020nni}. Conversely, starting with three-dimensional Einstein gravity, properties of a putative ensemble dual have been explored~\cite{Cotler:2020ugk}; see also \cite{Perez:2020klz, Maxfield:2020ale}.

An ensemble of theories satisfying Eqs.~\eqref{uninu} and \eqref{hawkout} may not exist in all cases where the RT prescription can be applied. If it does not, then the state paradox remains unresolved. For example, type IIB supergravity on AdS$_5\times \mathbf{S}^5$ is dual to a specific CFT~\cite{Mal97}, and no other boundary theories are presently known that have the same bulk as a coarse-grained description. If none exist, the gravitational path integral may still be expected to compute quantities that {\em would} be self-averaging if an ensemble {\em did} exist~\cite{Saad:2019lba}. It would determine the entropy $S(\rho_{\rm out})$ but not the state $\rho_{\rm out}$.

\paragraph{Outline} In this paper, we consider several distinct settings in which the state paradox appears, and we discuss its possible resolution by gauge/ensemble duality in each case. Multiple versions of the RT prescription will apply, and we will clarify their relation.

In Sec.~\ref{sh}, we use the RT prescription in an AdS$_d$ bulk spacetime to derive the Page curve in the dual CFT$_{d-1}$~\cite{Bousso:2019ykv}. The setting is distinct from that of of Refs.~\cite{Pen19,AEMM} in that there is no external bath or auxiliary system. The radiation remains in an AdS bulk and appears only on the right side of Eq.~\eqref{rtsimple}. The left hand side corresponds to the entropy of the CFT dual, for which a Page curve is obtained. The state paradox then arises in the CFT. The CFT entropy can also be computed as the von Neumann entropy of a CFT state constructed by applying the standard AdS/CFT extrapolate dictionary to the bulk. With this method, one finds that the CFT entropy should grow monotonically. These results are consistent only if the CFT is actually an ensemble of CFTs.

In Sec.~\ref{shb}, we turn to the setting of Refs.~\cite{Pen19,AEMM}. The gravitating spacetime is coupled to an auxiliary system without gravity, into which the Hawking radiation escapes. The entropy of the auxiliary system is computed using RT. The RT prescription must first be extended so that it applies to auxiliary systems; this was initially viewed as a weak link in the analysis~\cite{AEMM}. We argue that the correct prescription is fully determined by consistency with the setting of Sec.~\ref{sh}. The radiation appears on both sides of Eq.~\eqref{rtsimple}, leading to the state paradox unless gravity/ensemble duality is invoked.

Several works~\cite{AMMZ,Rozali:2019day,Chen:2019uhq,Almheiri:2019psy,Sully:2020pza,Bak:2020enw} have computed the Page curve using the entanglement wedge in a ``doubly holographic'' dual. The fundamental object in this case is the auxiliary system containing the radiation: a ``Boundary'' conformal field theory or BCFT (in the sense of Refs.~\cite{Car04,AffLud91}), with an apparently different RT prescription~\cite{Tak11,FujTak11}. The state paradox is somewhat obscured in this approach. To exhibit it, we deconstruct the RT prescription for BCFTs as a repeated application of the original RT prescription. 

For the sake of clarity, we first develop an RT prescription for a doubly holographic setting {\em without} auxiliary system, in Sec.~\ref{dh}. In this case, the fundamental object is a regular CFT$_{d-1}$ dual to an AdS$_d$ bulk. The bulk matter sector is assumed to consist of a holographic CFT$_d$ coupled to gravity. Then there exists a second holographic dual with $d+1$ dimensions. The original RT prescription computes the von Neumann entropy of a CFT$_{d-1}$ region as the generalized entropy of its entanglement wedge in the AdS$_d$ bulk. An adaptation of the RT prescription to braneworlds~\cite{Emp06,MyePou13} can be used to compute generalized entropy in the AdS$_d$ bulk using the $d+1$ dimensional bulk. We show that these steps can be combined into a one-step ``squared RT'' prescription for computing CFT$_{d-1}$ entropy from a ``squared entanglement wedge,'' EW$^2$, in the $d+1$ dimensional bulk. 

In Sec.~\ref{dhb}, we combine the settings of the previous two sections. We consider a doubly holographic CFT$_{d-1}$, coupled to a (singly) holographic CFT$_d$ that plays the role of the auxiliary system of Sec.~\ref{shb}. In the second holographic dual, the CFT$_d$ is part of the conformal boundary of the $d+1$ dimensional bulk. Like in Sec.~\ref{dh}, we show that the RT prescriptions for each holographic layer can be combined into a (one-step) squared RT prescription that uses the $d+1$ bulk to compute the von Neumann entropy of any union of subregions of the above top-level CFT$_{d-1}$ and CFT$_d$. %The boundary system in Sec.~\ref{dhb} can also be viewed as a ``Boundary" CFT (BCFT). 
Our squared RT prescription agrees with the known RT prescription for BCFTs~\cite{Tak11,FujTak11}. 

It follows that~\cite{Tak11,FujTak11} can be deconstructed as two applications of the RT prescription. This allows us to shed light on a number of puzzling features in Refs.~\cite{AMMZ,Rozali:2019day,Chen:2019uhq,Almheiri:2019psy,Sully:2020pza,Bak:2020enw}. We find that the state paradox arises at the first step, for the Hawking radiation that has escaped to the ``auxiliary'' CFT$_d$. At this level the paradox can be resolved by replacing (at least) the CFT$_{d-1}$ with an ensemble of such theories. 

The second level of holography furnishes a bulk dual of the original state paradox. The RT prescription for braneworlds computes the entropy of subregions of the first holographic dual, in terms of bulk quantities in the second dual. Choosing the subregion to be just the radiation region, this reproduces Hawking's rising curve; choosing it to include the island as well, one again obtains the Page curve. 

Finally, we observe that when the entropy of a top level CFT$_d$ region is computed directly using the squared RT prescription~\cite{Tak11,FujTak11}, no paradox is manifest, because the $d+1$ bulk dual does not contain the radiation.

\paragraph{Discussion} The discovery of entanglement islands~\cite{Pen19,AEMM} provides evidence for unitarity, independently of AdS/CFT. It marks a new era in which the Page curve can be derived from gravitational physics directly. It provides independent evidence for unitary evolution. However, it does not resolve the critical question of how the information gets out.

If we insist that information is preserved when a black hole evaporates, then effective field theory or General Relativity must break down substantially, at or outside of the horizon~\cite{AMPS}, at late times but while the horizon is still weakly curved. This formulation of the information paradox is called the firewall paradox. 

The firewall argument suggests that Hawking's ``mistake'' was the perfectly reasonable assumption that the horizon of a large old black hole is smooth. The AdS/CFT correspondence can be used to strengthen this argument~\cite{MarPol13}, but it has shed no light on the bulk dynamics that would produce a firewall. A number of interesting proposals attempt to reconcile unitarity with a smooth horizon; see Refs.~\cite{Almheiri:2013hfa,Harlow:2014yka} for a critical review. These proposals remain incomplete, and they appear to necessitate an element of nonlinearity that conflicts with the principles of quantum mechanics no less than information loss would~\cite{Bou12c,Bou13a,Bou13b}.

The bulk path integral derivation of the Page curve has been interpreted as a resolution of the firewall paradox~\cite{Pen19,AMMZ,Almheiri:2019qdq,Almheiri:2020cfm}. This seems plausible, since the bulk geometry involved in the calculation of the Page curve (top of Fig.~\ref{fig:intro}) has a manifestly smooth horizon. However, this picture just trades the firewall paradox for the state paradox~\cite{Bousso:2019ykv}. Then the question becomes how the state paradox is resolved. We see two possibilities.

Suppose that the state paradox is resolved by gravity/ensemble duality. The firewall argument~\cite{AMPS} does not apply to the ensemble averaged state, since its evolution is not unitary. Therefore it is consistent for the horizon to be smooth. However, fundamentally it makes no sense for Nature to be described by an ensemble of unitary theories; we can just measure the couplings and then work with the one correct theory. Moreover, we do expect the unique theory describing black hole formation and evaporation---the one that applies to an experiment conducted in a lab---to preserve information. The ensemble will be useful only for computing self-averaging quantities of the correct unitary theory, since these are the same in each theory; these evidently include the entropy, but not the final state. Hence the true S-matrix must be computed from a single unitary theory, not from an ensemble. In this theory, the firewall argument still applies.

If instead there is no gravity/ensemble duality (for example, in settings where no suitable ensemble exists, or where the gravity path integral cannot be rigorously defined), then the bulk path integral (or the saddlepoint approximation to it) would have to be viewed as an uncontrolled approximation. The path integral succeeds at computing certain quantities of a single unitary boundary theory (like the entropy) but not others (like the details of the late time state). Then there is no reason to trust the smooth geometry that appears in the input to the RT calculation, any more than we should trust the large entropy of the Hawking radiation that is manifest at this step. If we believe the output of the RT calculation---the Page curve---, then the firewall paradox precludes a smooth horizon.

\paragraph{Notation and Conventions} A subscript on a geometric object generally indicates not {\em its} dimension, but the dimension of the (physical) spacetime in which the object is naturally defined. For example, $M_d$ will denote a $d$-dimensional spacetime, $R_d$ a $d-1$ dimensional spatial region in $M_d$, and $\gamma_d$ a $d-2$ dimensional extremal surface in $M_d$. It is often useful to conformally rescale a manifold $M_d$ so that a boundary can be added to it; the result is called an unphysical manifold or Penrose diagram, $\tilde M_d\supset \partial\tilde M_d$. Note that the boundary of the physical manifold, $\partial M_d\subset \partial \tilde M_d$, need not be empty; it consists of braneworlds or end-of-the-world (EOW) branes.

In this paper, $\braket{x}$ always denotes the ensemble average of $x$ in the sense of Eq.~\eqref{average}. Angular brackets never denote a quantum expectation value.

The term {\em Boundary Conformal Field Theory} (BCFT) refers to the fact that such a theory lives on a manifold {\em with} Boundary, not to the fact that it lives on the conformal boundary {\em of} some AdS spacetime. We will capitalize ``Boundary'' whenever it is used in the sense of a BCFT. For example, ``boundary entropy" might refer to the von Neumann entropy of a CFT region on the conformal boundary $M_{d-1}=\partial \tilde M_d$, whereas ``Boundary entropy" is a specific BCFT parameter defined by Cardy~\cite{Car04}.

Throughout this paper we assume $d>2$ for convenience. The case $d=2$ would frequently require a special treatment; see for example Eq.~\eqref{gdd}. This would clutter the presentation. However, the qualitative aspects of our analysis apply in $d=2$, and hence to the many recent works that studied entanglement islands in JT gravity and other two-dimensional models, such as Refs.~\cite{AEMM,AMMZ}. Related to this choice, in examples involving braneworlds we only consider induced gravity on the brane (i.e., the localized graviton due to embedding of the brane in AdS~\cite{RS1}). We never add an additional gravitational action on the brane, because in $d>2$ this is not necessary.

\section{Gravity/Ensemble Duality Without a Bath}
\label{sh}

In this section, we exhibit a version of the state paradox in which only the standard RT prescription is needed~\cite{Bousso:2019ykv}. There is no auxiliary system or bath, and there is only one layer of holography.

\subsection{General Setup}
\label{shset}

\begin{figure}
    \centering
    \includegraphics[width=0.8\textwidth]{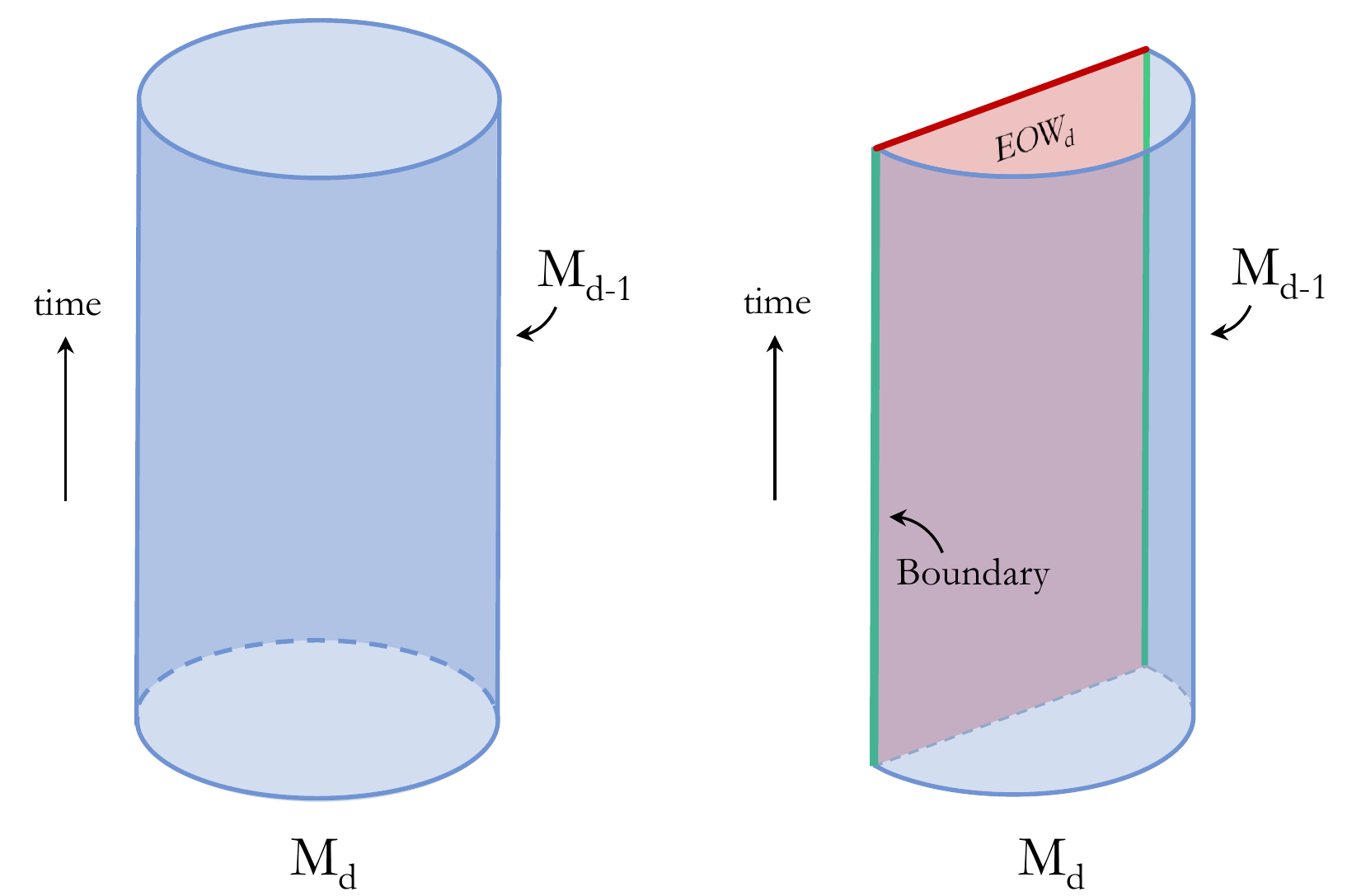}
    \caption{Examples of holographic duality. {\em Left:} The solid bulk $M_d$ is dual to a holographic CFT$_{d-1}$ on $M_{d-1}$ (blue boundary). {\em Right:} In this example, $M_{d-1}$ is a manifold with boundary, so the boundary theory is a BCFT$_{d-1}$ and $M_d$ contains an end of the world brane EOW$_d$. (Despite the appearance of a BCFT this is a ``singly holographic'' example. In Sections~\ref{dh} and $\ref{dhb}$ we will consider a doubly holographic setting where the EOW$_d$ is a braneworld that localizes gravity and contains a holographic CFT$_d$.)}
    \label{fig:2.1}
\end{figure}

Consider a $d-1$ dimensional\footnote{For consistency with the later sections on double holography, we deviate here from the usual convention of using $d$ for the boundary spacetime dimension.} holographic conformal field theory CFT$_{d-1}$ with central charge $c_{d-1}$, living on a manifold $M_{d-1}$; see Fig. \ref{fig:2.1}. Its bulk dual will be an asymptotically AdS$_d$ spacetime $M_d$,%
\footnote{In general this spacetime can contain additional factors, e.g.\ AdS$_d\times \mathbf{S}^{d'}$, so it need not actually be $d$-dimensional. In order to keep the discussion simple, we will assume that it is; generalizations are straightforward.}
such that the unphysical spacetime (or Penrose diagram) conformally related to $M_d$~\cite{Wald} is
\begin{equation}
    \tilde M_d = M_d \cup M_{d-1}~;
\end{equation}
thus $M_{d-1}$ is the conformal boundary of $M_d$.  The AdS$_d$ curvature length $L_d$ is related to the central charge by
\begin{equation}
\frac{L_d^{d-2}}{G_d}\sim c_{d-1}~,
\end{equation}
where $G_d=\ell_d^{d-2}$ is Newton's constant in the $d$-dimensional bulk. 

We shall denote a standard holographic duality of this type as follows:
\begin{equation}
    M_{d-1} \longrightarrow M_d~,
\end{equation}
where the arrow reminds us that in general, this duality is not truly an equivalence. Rather, the lower dimensional field theory without gravity can be viewed as the nonperturbative completion of the bulk theory.

Note that $M_{d-1}$ may itself have a boundary, as in Fig. \ref{fig:2.1}.
The spacetime $M_d$ may also be a manifold with boundary~\cite{Wald}, commonly referred to as an ``end of the world brane'' or EOW:
\begin{equation}
\mbox{EOW}_d \equiv \partial M_d~.
\end{equation}
In particular, if $M_{d-1}$ is a manifold with boundary, then the CFT$_{d-1}$ is a ``Boundary conformal field theory'' (BCFT),\footnote{See the notation section at the end of the introduction.} and the bulk $M_d$ will contain an EOW$_d\neq \varnothing$ anchored on the Boundary $\partial M_{d-1}$~\cite{Tak11,FujTak11}. 
%We refer to $\partial M_{d-1}$ as the ``Boundary" with a capital ``B." 
An EOW can also exist in settings where the Lorentzian CFT$_{d-1}$ has no Boundary~\cite{Kourkoulou:2017zaj,Cooper:2018cmb}. They must be included in the gravitational path integral.

\subsection{Ryu-Takayanagi Prescription}
\label{shrt}

We now formulate the holographic prescription for computing the von Neumann entropy of a boundary region from bulk quantities. This was first proposed by Ryu and Takayanagi~\cite{RyuTak06,RyuTak06b} for stationary states. It was generalized to the time-dependent case by Hubeny, Rangamani, and Takayanagi~\cite{HubRan07}, and to the BCFT case by Takayanagi and collaborators~\cite{Tak11,FujTak11}. A quantum-corrected prescription was first proposed by Faulkner, Lewkowycz and Maldacena~\cite{FauLew13}. It was extended to all orders by Engelhardt and Wall~\cite{EngWal14}, whose elegant formulation highlights the central role of generalized entropy. 
\begin{figure}
    \centering
    \includegraphics[width=0.85\textwidth]{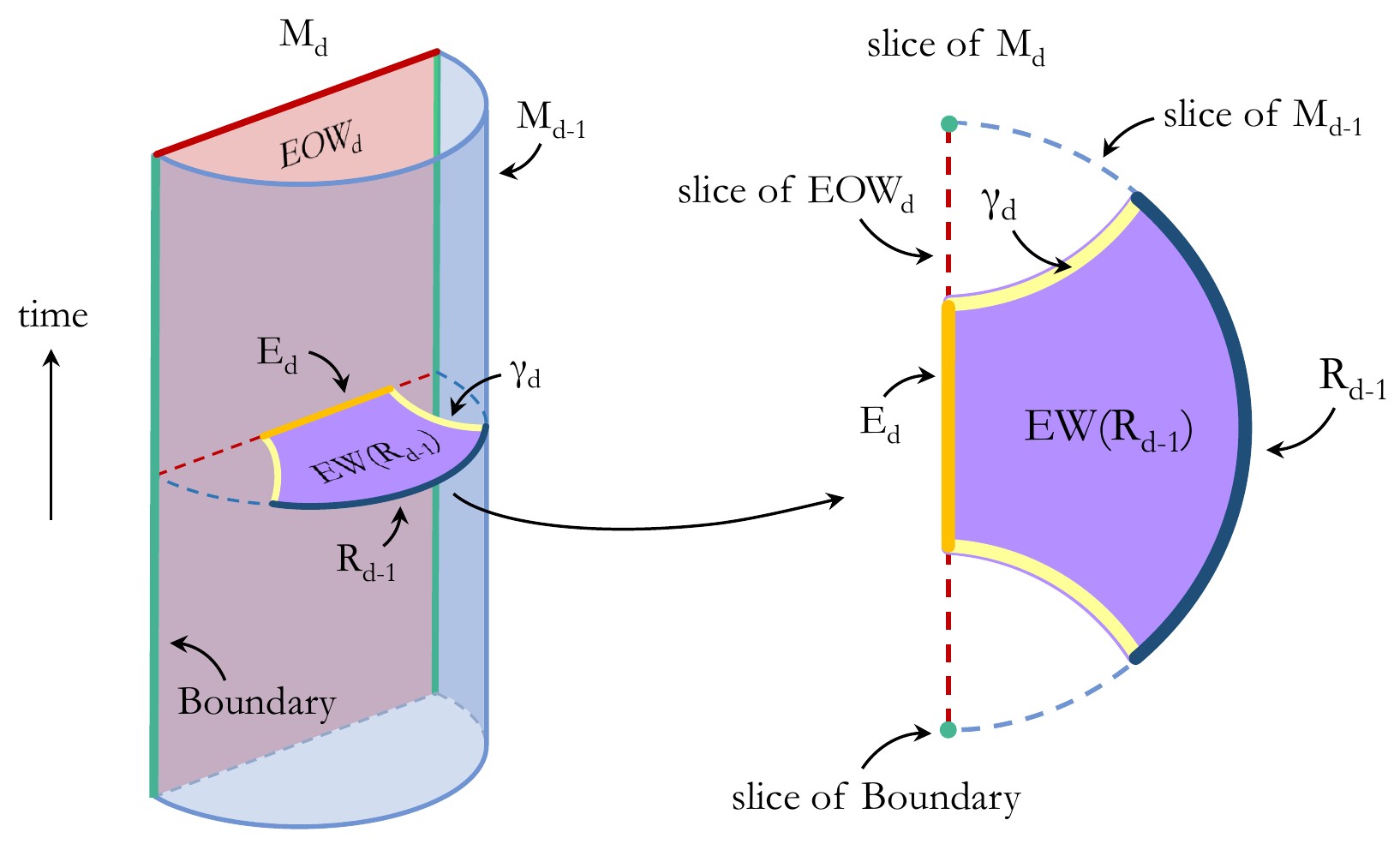}
    \caption{RT prescription, applied in the setting shown on the right of Fig.~\ref{fig:2.1}. The entropy of the boundary region $R_{d-1}$ is given by the generalized entropy of its entanglement wedge EW$(R_{d-1})$. $\gamma_d$ is the quantum extremal surface.} 
    \label{fig:2.2}
\end{figure}
This final formulation is essential for the existence of islands,\footnote{Because the empty surface always has less classical area than the boundary of an island, area minimization cannot lead to an island. It is vital that the generalized entropy is minimized.} and it is the only one we will review here. We will refer to it as the RT prescription for short, with apologies to all others involved in its development. We aim to make it clear throughout this paper that islands are part and parcel of this prescription. They do not constitute a new ingredient, but a long-overlooked consequence. The recent recognition of their existence~\cite{Pen19,AEMM} has been profoundly impactful.

Let $R_{d-1}\subset M_{d-1}$ be an achronal region (see Fig.~\ref{fig:2.2}).\footnote{An achronal region is a submanifold of codimension 1 (in the spacetime) which contains no two points connected by a timelike curve.} We can think of $R_{d-1}$ as a subregion at some instant of time, to which the CFT$_{d-1}$ state may be restricted. The von Neumann entropy $S(R_{d-1})$ of the restricted CFT$_{d-1}$ state is given by the generalized entropy of its entanglement wedge,
\begin{equation}
  S(R_{d-1}) = S_{\rm gen}[\mbox{EW}(R_{d-1})]~.
  \label{qhrt}
\end{equation}
The generalized entropy $S_{\rm gen}(X_d)$~\cite{Bek72} of an arbitrary achronal region $X_d\subset M_d$ is the sum of its gravitational entropy and the von Neumann entropy $S$ of the quantum fields in the region $X_d$:
\begin{equation}
  S_{\rm gen}(X_d) =\frac{{\cal A}(\partial X_d)}{4G_d} +S(X_d)~.
  \label{sgendef}
\end{equation}
Here ${\cal A}(\partial X_d)$ is the area of the boundary of $X_d$ in $M_d$, and $G_d$ is Newton's constant in $M_d$.

The \emph{entanglement wedge} EW$(R_{d-1})$ is an achronal region $X_d$\footnote{One can also define EW$(R)$ to be the ($d$-dimensional) domain of dependence of this region. Since all Cauchy slices of the domain of dependence have the same generalized entropy, we will use these definitions interchangeably.} in $M_d$, that satisfies the following conditions:
\begin{enumerate}
\item \textit{Homology}: $\partial X_d=\gamma_d\cup R_{d-1}\cup E_d$, where $\gamma_d\subset M_d-\mbox{EOW}_d$, and $E_d\subset$ EOW$_d$.\footnote{Strictly, this is a statement about the image of $X_d$ in the unphysical spacetime $\tilde M_d$.} See Fig. \ref{fig:2.2}.
\item \textit{Stationarity}: $S_{\rm gen}(X_d)$ is stationary under variations of $\gamma_d$.
\item \textit{Minimality}: $X_d$ is has the smallest $S_{\rm gen}$ among all regions with the above properties.
\end{enumerate}

A surface $\gamma_d$ satisfying the homology constraint (1) and the stationarity condition (2) is called \emph{quantum extremal}\footnote{This is conventional. ``Quantum stationary'' would be more appropriate terminology, as the generalized entropy can be both increased and decreased at second order by suitable deformations.} with respect to $R_{d-1}$. If the minimality condition (3) is also satisfied, then $\gamma_d$ is called the \emph{RT surface} of $R_{d-1}$. Note that $\gamma_d$ may be the empty set; for example, see Fig.~\ref{fig:2.3} below. Also, $\gamma_d$ may contain disconnected components that end neither on $R_{d-1}$ nor on $E$; for example, see Fig.~\ref{fig:2.4} below.

\subsection{Simple Boundary Unitarity from a Semiclassical Bulk}
\label{simple}

It was recently shown that the RT prescription applied to semiclassical bulk evolution yields an entropy consistent with boundary unitarity~\cite{Pen19,AEMM}, for Hawking radiation extracted into an auxiliary system. This argument requires an extension of the RT prescription that includes auxiliary systems. We will show in Sec.~\ref{shbrt} that this extension is uniquely determined by physical considerations. 

However, the main result of Refs.~\cite{Pen19,AEMM} can be obtained without involving an auxiliary system, using only the standard RT prescription, Eq.~\eqref{qhrt}. Here we summarize this argument; further details are discussed in Ref.~\cite{Bousso:2019ykv}.

\begin{figure}
    \centering
    \includegraphics[width=\textwidth]{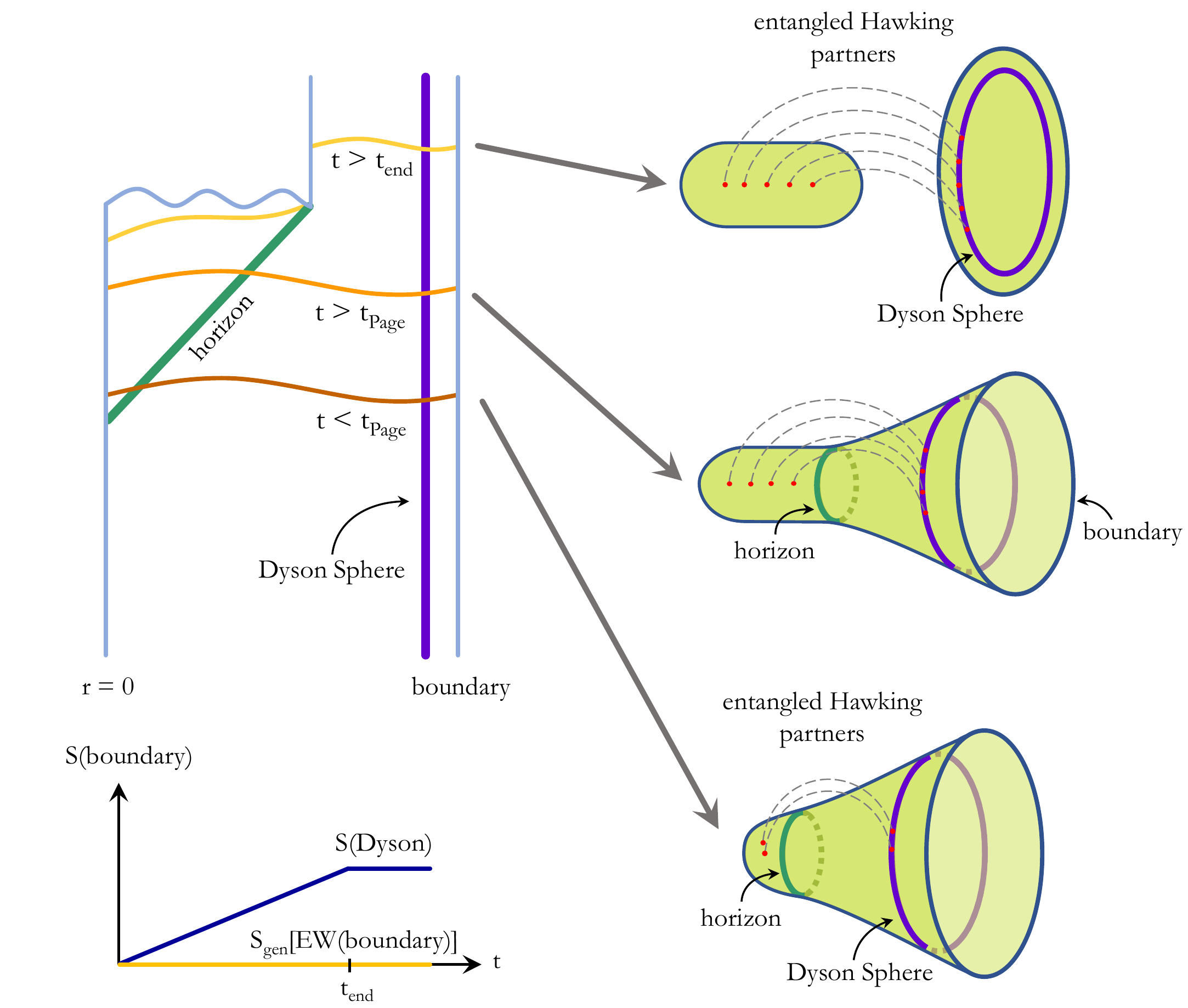}
    \caption{Hawking radiation is absorbed by a distant Dyson sphere near the boundary. In Hawking's semiclassical analysis, the Dyson sphere entropy will grow monotonically. The quantum state on the global bulk slices shown is pure. Each global slice is the entanglement wedge of its respective boundary slices. Thus the RT prescription implies that the entropy of the global boundary vanishes, as required by CFT unitarity. However, at late times the extrapolate dictionary demands that $S(\mbox{boundary})=S(\mbox{Dyson})$. This contradiction is the state paradox.}
    \label{fig:2.3}
\end{figure}

Consider a CFT$_{d-1}$ on $M_{d-1}=\mathbf{S}^{d-2}\times \mathbf{R}$. In the vacuum, the gravity dual $M_d$ would be global AdS$_d$. However, we shall take $M_d$ to be a black hole formed from collapse of matter in a pure quantum state. The black hole is surrounded by a distant detector sphere (``Dyson sphere"), initially in some pure reference state. By the extrapolate dictionary, the initial boundary state must be pure. As the black hole evaporates, the Dyson sphere absorbs all of the Hawking radiation (see Fig. \ref{fig:2.3}). 

Let $\Sigma_{d-1}(t)$ be a family of Cauchy surfaces (time slices) of the boundary $M_{d-1}$. Each such slice will be a sphere $\mathbf{S}^{d-2}$. Three slices are shown in Fig. \ref{fig:2.3}. We will apply the RT prescription to every slice, but first it will be useful to make some further definitions. Let $\Sigma_d(t)$ be a Cauchy surface of $M_d(t)$ bounded by $\Sigma_{d-1}(t)$. (In $\tilde M_d$, $\Sigma_{d-1}=\partial \Sigma_d$.) For boundary slices that lie in the future of the endpoint of the evaporation process, we define $\Sigma_d(t)$ to include a disconnected component, a Cauchy slice of the black hole interior (see Fig. \ref{fig:2.3}, yellow slice at top). This can be chosen far enough from the singularity so that semiclassical gravity is applicable everywhere but in the neighborhood of the endpoint~\cite{Lowe:1995ac}.

The key observation is that the entanglement wedge is the entire bulk:
\begin{equation}
    \Sigma_d(t) = \mbox{EW}[\Sigma_{d-1}(t)]~,
\end{equation}
for all $t$. To see this, note that the homology condition is satisfied, with $\gamma_d=\varnothing$. The stationarity condition is satisfied because no variations of $\gamma_d$ exist. The minimality condition is satisfied because
\begin{equation}
    S_{\rm gen}[\Sigma_d(t)] = 0 
    \label{sgenzero}
\end{equation}
for all $t$, and the generalized entropy cannot be negative. 

(Strictly, one could question all three of these statements due to the breakdown of the semiclassical description at the evaporation endpoint. We assume that this small region does not contribute significant effects that invalidate our treatment of the post-evaporation entanglement wedge. In any case, the essence of our discussion requires us only to go past the Page time, but not close to or beyond the endpoint of evaporation.)

It is important to understand why Eq.~\eqref{sgenzero} holds. The area term in Eq.~\eqref{sgendef} vanishes since $\gamma_d=\varnothing$. The von Neumann entropy of the matter fields vanishes because the initial bulk state is pure, and the semiclassical bulk evolution of the {\em global} bulk state is unitary. (Information is lost to an observer outside the black hole in this description~\cite{Haw74}, but globally the state remains pure. The interior Hawking partners and the exterior Hawking radiation together form a pure state, the vacuum at the horizon.) 

By Eq.~\eqref{qhrt}, it follows that 
\begin{equation}
    S[\Sigma_{d-1}(t)]=0
\end{equation}
for all $t$. The RT prescription ``predicts" that the entropy of the boundary theory vanishes at all times. Of course, this is exactly what is expected from the unitarity of the boundary CFT$_{d-1}$. But it is remarkable that this result is reproduced by performing a semiclassical analysis in the bulk---the same calculation that led Hawking to conclude that information is lost to bulk observers outside the black hole. This fact was perhaps not widely appreciated prior to the recent work~\cite{Pen19,AEMM} that derives the entire Page curve, even though it has the same import and is simpler to obtain.

\subsection{Island and Page Curve}
\label{refined}

\begin{figure}
    \centering
    \includegraphics[width=.9\textwidth]{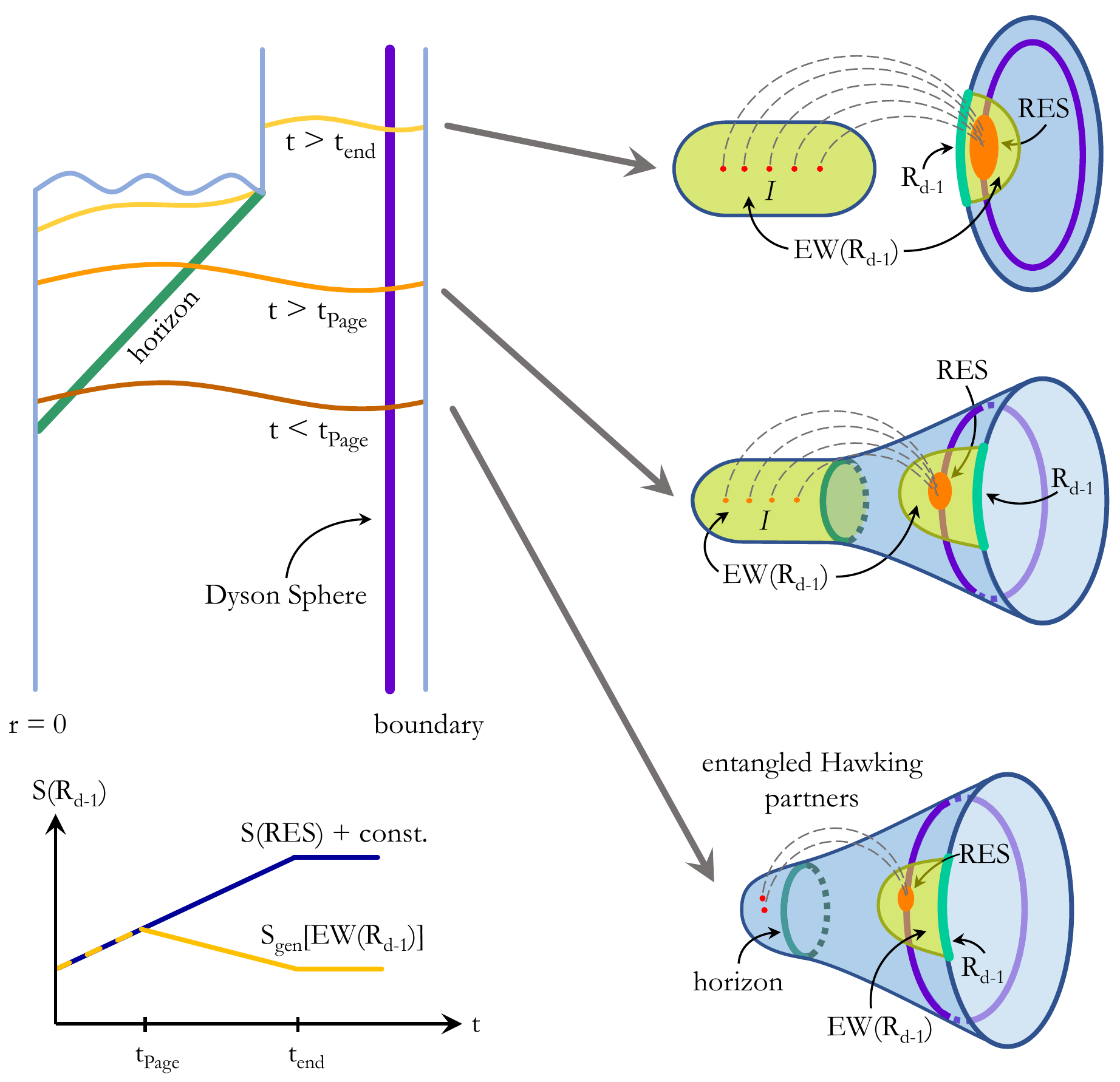}
    \caption{Compared to Fig.~\ref{fig:2.3}, the Hawking radiation is collected in a localized reservoir on the Dyson sphere. The RT prescription is applied to a nearby boundary region $R_{d-1}$. The entanglement wedge EW$(R_{d-1})$ is shown in light green. After the Page time, it contains a disconnected island $I$, the black hole interior, because this choice minimizes the generalized entropy. This yields the Page curve for $S(R_{d-1})$. However, the extrapolate dictionary would yield Hawking's curve; this is the state paradox.}
    \label{fig:2.4}
\end{figure}

The previous subsection explained how the RT prescription yields the vanishing global boundary entropy consistent with unitarity, despite using Hawking's semiclassical evolution in the bulk. In this subsection, we introduce a refined scenario, such that the RT prescription yields the Page curve for two complementary subsystems, the Hawking radiation and the remaining black hole. In order to implement this without introducing an external bath or auxiliary system, any absorbed Hawking radiation is immediately transferred to a localized reservoir $\mbox{RES}$ taking up a small solid angle on the Dyson sphere, without loss of quantum coherence (see Fig. \ref{fig:2.4})~\cite{Bousso:2019ykv}.

Gravitational backreaction in the asymptotic region can be kept arbitrarily small, so the shape of any stationary surface anchored to a small boundary region $R_{d-1}$ will be similar to that in the vacuum. We take $R_{d-1}(t)\subset \Sigma_{d-1}(t)$ to be at the same angular position as the reservoir, and just large enough so that EW$[R_{d-1}(t)]$ will barely contain the reservoir (see Fig. \ref{fig:2.4}). Before the Page time, the entanglement wedge has only one connected component, and we find
\begin{equation}
S[R_{d-1}(t)] = S_{\rm gen}[\mbox{EW}(R_{d-1}(t))]=\frac{\mathcal{A}[\gamma_d^{\rm conn}]}{4G} + S_{\rm RES}(t) ~.
\end{equation}
The superscript refers to the fact that $\gamma_d=\gamma_d^{\rm conn}$ is connected to $R_{d-1}$ before the Page time. By moving around ballast on the Dyson sphere, one can arrange for the asymptotic geometry in an open neighborhood of $\gamma_d^{\rm conn}$, and hence for $\mathcal{A}[\gamma_d^{\rm conn}]$ to remain fixed~\cite{Bousso:2019ykv}. The entropy of the reservoir $S_{\rm RES}(t)$, however, increases as more radiation arrives. This yields the rising part of the Page curve shown in Fig. \ref{fig:2.4}. 

The entropy of the Dyson sphere, and of $S_{\rm RES}$ in particular, increases monotonically even after the Page time. Its state is always purified by the ``Hawking partners'' in the black hole interior. Inclusion of the black hole interior in the entanglement wedge will entirely wipe out the contribution $S_{\rm RES}$ to $S_{\rm gen}(R_{d-1})$ at a cost of increasing the area term by the area of the black hole. This preserves the homology condition, since it merely adds an extra component to $\gamma_d$. By its very definition, this choice becomes favorable at the Page time, when the black hole and radiation entropy are equal.

After the Page time, the minimality condition thus requires that EW$(R_{d-1})$ contains a second, disconnected component $I$ (see Fig. \ref{fig:2.4}). This is called an island, in the terminology of Ref.~\cite{AMMZ}. The island is the black hole interior, bounded by a disconnected component $\gamma_d^{\rm island}(t)$ that nearly coincides with the horizon.\footnote{The precise location of $\gamma_d^{\rm island}$ is determined by the stationarity condition. It sits about a Planck length inside the horizon. Temporally, $\gamma_d^{\rm island}(t)$ is located at $t-t_{\rm scr}$, where $t_{\rm scr}\sim \beta \ln\left(\mathcal{A}[\gamma_d^{\rm island}]/4G\right)$ and $\beta$ is the inverse temperature of the black hole~\cite{Pen19,AEMM}.} The interior of $\gamma_d^{\rm island}$ purifies the Hawking radiation, so the entropy of the reservoir no longer contributes, and $S_{\rm gen}[\mbox{EW}(R_{d-1})]$ is given just by the area of the RT surface $\gamma_d = \gamma_d^{\rm conn} \cup \gamma_d^{\rm island}$:
\begin{equation}
S_{\rm gen}[\mbox{EW}(R_{d-1})] = \frac{\mathcal{A}[\gamma_d^{\rm conn}]}{4G}  + \frac{\mathcal{A}[\gamma_d^{\rm island}(t)]}{4G} ~.
\end{equation}
The first term remains constant. But $\gamma_d^{\rm island}(t)$ shrinks with the black hole horizon as the black hole evaporates, yielding the decreasing part of the Page curve. 

Thus, in the refined scenario, the RT prescription (i.e., a bulk path integral that computes the entropy) yields the Page curve for the boundary region $R_{d-1}$. It rises during the first half of the evaporation process, then decreases. Again, this is consistent with our expectations from boundary unitarity. Entanglement wedge complementarity is manifest in the present setting, so a Page curve is also obtained for the complementary boundary region $\bar R_{d-1}$.

\subsection{State Paradox and Ensemble Interpretation}
\label{ensemble}

The large entropy of the Dyson sphere at late times leads to the state paradox. After the evaporation is complete, all of the (conserved) mass is in the Dyson sphere. The standard AdS/CFT dictionary can be used to construct the boundary state from the mixed state of the Dyson sphere~\cite{HKLL}. It dictates that the boundary state must have the same entropy as the Dyson sphere. Energetic arguments preclude purification of this state by some nonlocal CFT excitations~\cite{Bousso:2019ykv}. The entropy of the CFT$_{d-1}$ should therefore grow monotonically throughout the evaporation process. But this contradicts both the RT result and the expected unitarity of the boundary theory.

We stress again that one cannot dismiss the large Dyson sphere entropy as an artifact of the semiclassical approximation, without discarding the entire RT calculation. If the reservoir RES did not have large entropy after the Page time, the black hole interior could not purify it. Then there would be no reason to include the island.

\begin{figure}
    \centering
    \includegraphics[width=\textwidth]{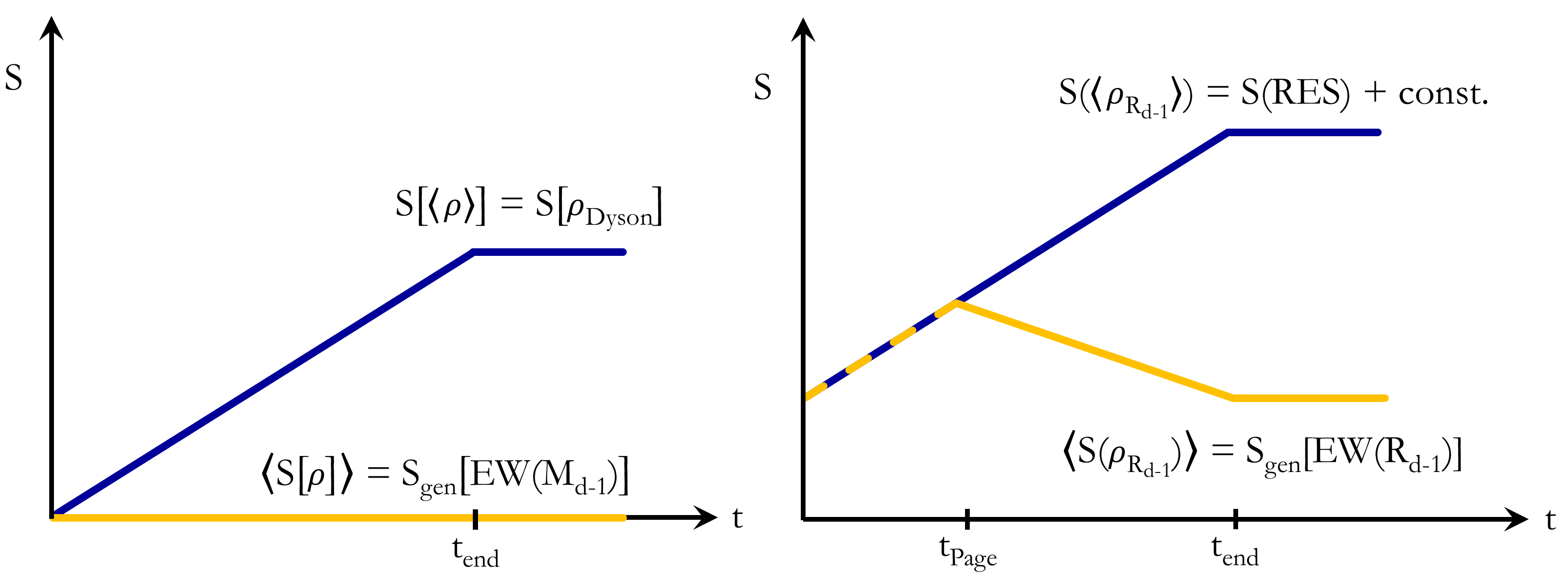}
    \caption{Here we assume the gravity/ensemble interpretation, in the examples studied in Sec.~\ref{simple} (left) and Sec.~\ref{refined} (right). This resolves the state paradox. The RT prescription (yellow) computes the ensemble averaged entropy $\braket{S}$ of the full boundary (left) or of $R_{d-1}$ (right). The extrapolate dictionary (blue) yields the average state of the ensemble, $\braket{\rho}$, in these regions.}
    \label{fig:2.5}
\end{figure}

In the setting of this section, the paradox does not arise for the state of the bulk radiation, but for the boundary state, since we are using the RT prescription to compute the entropy of the latter. A resolution of the state paradox can then be obtained by assuming that the boundary CFT is an \emph{ensemble} of unitary theories, and that the boundary quantities computed using the bulk are ensemble averages (see Fig. \ref{fig:2.5}). This proposal is consistent both with the smallness of $S[R_{d-1}(t)]$ and the fact that the reservoir contains a mixed state, for $t>t_{\rm Page}$. Since each member of the ensemble is unitary, $S(R_{d-1})$ must follow the Page curve in each theory. Hence the ensemble average of $S(R_{d-1})$ also follows the Page curve.

But the state of $R_{d-1}$ need not be self-averaging. Each member of the ensemble predicts a pure out-state, but this need not be the {\em same} pure out-state in each theory. Hence the ensemble average of the out-state is a mixed state whose entropy can continue to grow after the Page time. Under the ensemble interpretation, the ensemble-averaged boundary state can be obtained by applying the standard AdS/CFT dictionary to the semiclassical bulk state.

The most explicit calculations of entanglement islands so far~\cite{AEMM} were done for the case where the bulk is JT gravity, which is indeed dual to a matrix ensemble. However, we stress that the above argument is unrelated to this observation. The state paradox should be viewed as independent evidence that the gravity path integral, if it is well defined, must be dual to an ensemble, even in settings where no suitable ensemble dual is currently known.

\section{Gravity/Ensemble Duality With a Bath}
\label{shb}

In this section, we turn to the settings studied by Penington~\cite{Pen19} and by Almheiri {\em et al.}~\cite{AEMM}. We will argue that the relevant RT prescription can be deduced from the standard one by requiring consistency with the analysis of the previous section. Finally, we will exhibit the state paradox and discuss its resolution by gravity/ensemble duality.

\begin{figure}
    \centering
    \includegraphics[width=\textwidth]{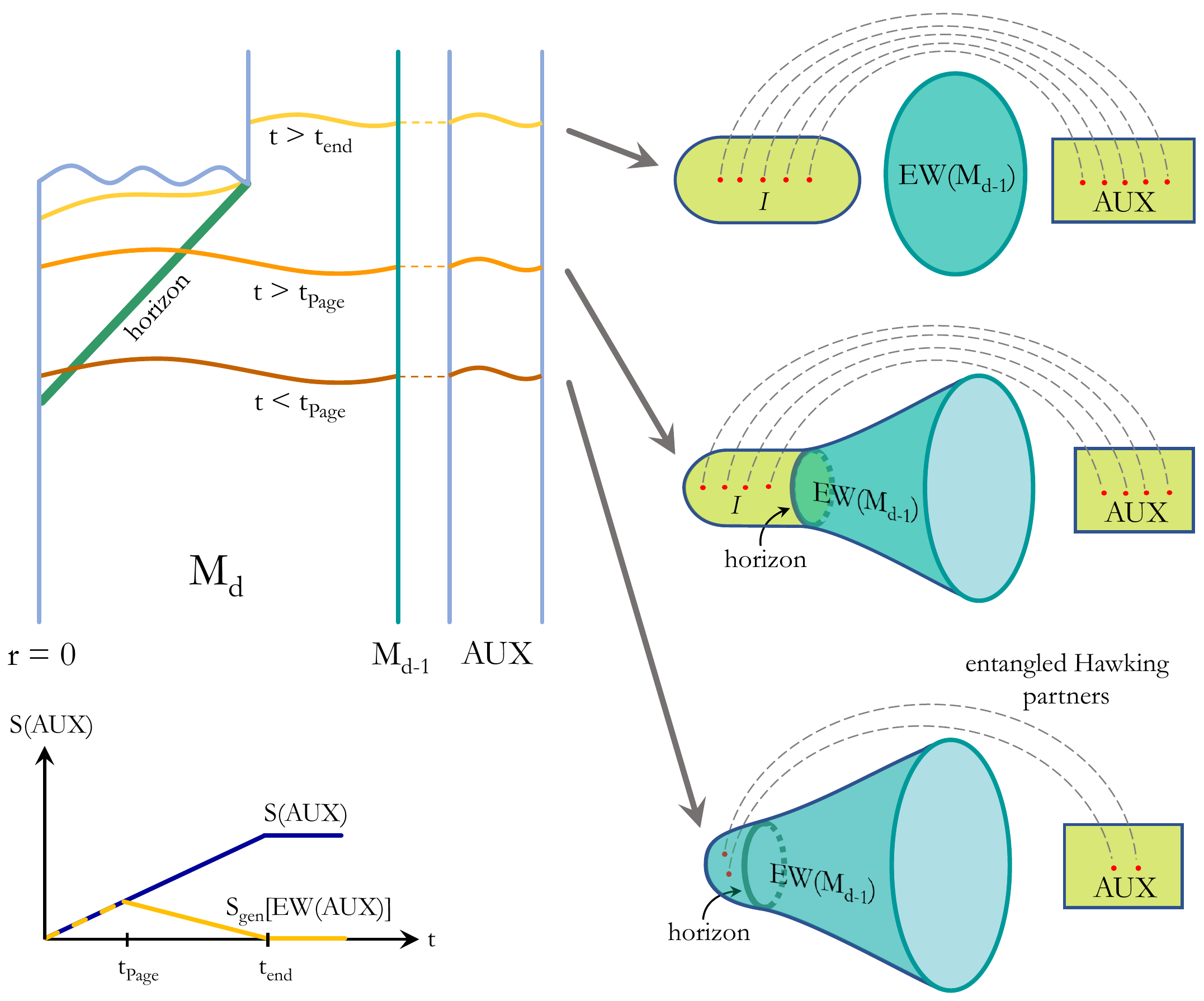}
    \caption{Hawking radiation escapes into an auxiliary system without gravity. The RT prescription can be applied to the boundary $M_{d-1}$, yielding the dark green entanglement wedge. A version of the RT prescription for AUX can be developed by requiring consistency with the analysis in Sec.~\ref{refined}. One finds that EW(AUX) (light green) includes AUX itself, and and after the Page time, it also the bulk region $I$ complementary to $EW(M_{d-1})$. The state paradox arises in AUX: the entropy must follow Hawking's rising curve for the island $I$ to be part of EW(AUX), but with $I$ included, RT yields the Page curve for AUX.}
    \label{fig:3}
\end{figure}

In contrast to Sec.~\ref{sh}, the Dyson sphere in AdS$_d$ is eliminated and replaced by an auxiliary (external) system AUX: a ``bath'' that couples to the boundary CFT$_{d-1}$ and absorbs the Hawking radiation (see Fig. \ref{fig:3}). Thus we study the holographic duality
\begin{equation}
   M_{d-1} \cup \mbox{AUX} \longrightarrow M_d \cup \mbox{AUX}  ~.
\end{equation}
In Ref.~\cite{AEMM}, AUX is a 1+1 dimensional CFT, and the black hole has two asymptotic regions. For definiteness, we will follow Penington~\cite{Pen19}, who considered the more physical setting of a black hole formed from collapse. The auxiliary system AUX will remain unspecified in this section.

We will begin by extending the RT prescription to include AUX, in Sec.~\ref{shbrt}. Previous efforts to extend the prescription required additional assumptions, such as entanglement wedge complementarity (defined below)~\cite{AEMM}. We present a novel argument that this extension is fully determined by the analogy between AUX and the reservoir RES in the previous section. EW complementarity is a consequence rather than an assumption of our argument.

In Sec.~\ref{shbip} we apply the RT prescription to black hole evaporation into AUX. This is just for completeness: we summarize Refs.~\cite{Pen19,AEMM} and restate the analysis in Sec.~\ref{refined} in this modified setting. The paradox identified in Ref.~\cite{Bousso:2019ykv} and reviewed in Sec.~\ref{ensemble} also has an analogue in this setting. In Sec.~\ref{shbap} we discuss this and its resolution if the bulk is dual to an ensemble of boundary theories.

\subsection{Ryu-Takayanagi Prescription With Auxiliary Systems}
\label{shbrt}

Consider a bipartite system consisting of a holographic CFT$_{d-1}$ in a region $R_{d-1}\subset M_{d-1}$ and an auxiliary system AUX, in some joint state. Suppose that there exists an RT-like prescription for computing the von Neumann entropy of this state. We shall take ``RT-like'' to mean that the prescription is of the form
\begin{equation}
  S(R_{d-1}\cup \mbox{AUX}) = S_{\rm gen}[\mbox{EW}(R_{d-1}\cup \mbox{AUX})]~.
  \label{island}
\end{equation}
We now determine the detailed formulation of the prescription from general considerations.

For any bipartite system consisting of a gravitating region $X_d$ and an auxiliary system AUX, we define the generalized entropy as 
\begin{equation}
  S_{\rm gen}(X_d\cup \mbox{AUX}) =\frac{{\cal A}(\partial X_d)}{4G_d} +S(X_d\cup \mbox{AUX})~.
 \label{sgenaux}
 \end{equation}
Given these definitions, the nontrivial content of the prescription we seek lies in how we define the entanglement wedge $\mbox{EW}(R_{d-1}\cup \mbox{AUX})$.

Entanglement wedge nesting, the property that the entanglement wedge cannot shrink if the boundary algebra is enlarged~\cite{Wall:2012uf}, implies that
\begin{equation}
\mbox{EW}(R_{d-1}\cup \mbox{AUX})\supset \mbox{EW}(R_{d-1})~. 
\label{nesting}
\end{equation} 
We next recall that the relative entropy between two boundary states, $S(\rho|\sigma)$, is the same as the relative entropy between the dual bulk states in the entanglement wedge~\cite{Jafferis:2015del}. This implies that bulk operators in the entanglement wedge (but not outside) can be implemented on the boundary~\cite{Almheiri:2014lwa,Jafferis:2015del,Dong:2016eik}. In particular, small deformations of the boundary state do not change the entanglement wedge. Taking $R_{d-1}\cup \mbox{AUX}$ as the boundary, consider a small deformation of the state in AUX. This can change the boundary relative entropy (in $R_{d-1}\cup \mbox{AUX}$), but it cannot change the bulk relative entropy in $\mbox{EW}(R_{d-1}\cup \mbox{AUX})$, unless we require that 
\begin{equation}
    \mbox{EW}(R_{d-1}\cup \mbox{AUX})\supset\mbox{AUX}.
    \label{auxaux}
\end{equation}
Therefore, AUX plays an interesting dual role: it appears both on the bulk and on the boundary side.

This does not yet fully determine the prescription. For example, Eqs.~\eqref{nesting} and \eqref{auxaux} would be consistent with the (wrong) proposal that $\mbox{EW}(R_{d-1}\cup \mbox{AUX})$ is given by $\mbox{EW}(R_{d-1})\cup \mbox{AUX}$. To see that this fails, we note that quantum information can be freely exchanged between AUX and $R_{d-1}$ by appropriate couplings. But consider the setting of Sec.~\ref{refined}. Recall that at the Page time, $EW(R_{d-1})$ has a phase transition: it now includes not only the portion connected to $R_{d-1}$, but also an island inside the black hole. Just after the Page time, let us couple the region $R_{d-1}$ to an AUX system that is initially in some pure reference state, and transfer some of the quantum information of the Hawking radiation into AUX. Then the phase transition is reversed; $EW(R_{d-1})$ loses the island. However, bulk operators in the island could be implemented on $R_{d-1}\cup \mbox{AUX}$ before the transfer, so this must still be true afterwards. Therefore, EW$(R_{d-1}\cup \mbox{AUX})$ cannot have changed.

This shows (at physics-level rigor) that an appropriate definition of the entanglement wedge must treat the bulk and AUX jointly, not separately, when minimizing the generalized entropy.  Hence we define
\begin{equation}
\mbox{EW}(R_{d-1}\cup \mbox{AUX}) \equiv X_d\cup \mbox{AUX}~,
\label{ewshb}
\end{equation}
where the spacetime region $X_d\subset M_d$ is chosen such that
\begin{enumerate}
\item $\partial X_d=\gamma_d\cup R_{d-1}\cup E_d$, where $\gamma_d\subset M_d-\mbox{EOW}_d$ and $E_d\subset$ EOW$_d$.
\item $S_{\rm gen}(X_d\cup \mbox{AUX})$ is stationary under variations of $\gamma_d$.
\item $X_d\cup \mbox{AUX}$ has the smallest $S_{\rm gen}$ among all regions $X_d$ with the above properties.
\end{enumerate}
We have included the possibility that $M_d$ has an EOW brane for generality, though none appears in the setup studied above. Note that the last term in Eq.~\eqref{sgenaux} would vanish in a case where $X_d$ and AUX separately have large von Neumann entropy but purify each other. Note also that AUX in the above formulas could represent one of several auxiliary systems, or equivalently, an arbitrary subalgebra of an auxiliary system.

The generalized RT prescription formulated above upholds entanglement wedge complementarity. Consider a pure quantum state for the complete system $M_{d-1}\cup$ AUX. On the boundary, purity implies $S_{d-1}(R_{d-1}\cup \mbox{AUX})=S_{d-1}(\bar R_{d-1})$, where $\bar R_{d-1}$ is complement of $R_{d-1}$ in $M_{d-1}$. Purity also implies $\gamma_d(M_{d-1}\, \cup\, \mbox{AUX})=\varnothing$. Hence $\mbox{EW}(M_{d-1}\cup \mbox{AUX})=M_d \cup \mbox{AUX}$.  The global bulk von Neumann entropy must also vanish: $S(M\cup \mbox{AUX})=0$. %whence $S_{\rm gen}[EW(CFT_d\cup AUX)]=S[EW(CFT_d\cup AUX)]=0$ by the RT prescription. 
This in turn implies that any two subsystems of $M\cup \mbox{AUX}$ must have equal von Neumann entropy. Therefore $\gamma_d(\bar R_{d-1}) = \gamma_d(R_{d-1}\cup \mbox{AUX})$, and hence 
\begin{equation}
\mbox{EW}(\bar R_{d-1}) = \overline{\mbox{EW}(R_{d-1} \cup \mbox{AUX})}~.
\label{complementarity}
\end{equation}

In the special case where AUX is a nongravitating system described by quantum field theory and $R_{d-1}=\varnothing$, Eq.~\eqref{island} reduces to the ``island formula'' of Ref.~\cite{AMMZ}, where it was derived using doubly holographic systems. The formula was already used implicitly by Penington~\cite{Pen19}. We have argued here that it emerges as a direct consequence of the standard RT prescription, when auxiliary systems are involved. We saw in Sec.~\ref{refined} that the standard RT prescription (without auxiliary systems) already required disconnected islands to be part of the entanglement wedge. The new aspect in the present discussion is not the possibility of an island, but the double role of AUX. This double role will make the state paradox particularly sharp.

\subsection{Island and Page Curve}
\label{shbip}

Returning to the specific setting of a black hole evaporating into AUX~\cite{Pen19}, we now examine the implications of Eq.~\eqref{island}. These follow immediately from the results of Sec.~\ref{refined}, upon substituting $R_{d-1}\to \mbox{AUX}$ and $\bar R_{d-1}\to M_{d-1}$. The entropy of each system follows a Page curve, as we will now verify.

Recall that $\Sigma_{d-1}(t)$ defines a foliation of the boundary $M_{d-1}$, and $\Sigma_d(t)$ are bulk Cauchy slices whose boundary is $\Sigma_{d-1}(t)$. Before the Page time, one finds that the entanglement wedge of the CFT$_{d-1}$ includes the entire bulk:
\begin{equation}
\mbox{EW}[\Sigma_{d-1}(t)] = \Sigma_d(t)~.
\end{equation}
Since $\gamma_d=\varnothing$ and hence $\mathcal{A}(\gamma_d)=0$,
\begin{equation}
S[\Sigma_{d-1}(t)] = S_{\rm gen}[\Sigma_d(t))]=S[\Sigma_d(t)]~.
\end{equation}
This grows with time, because $\Sigma_d(t)$ contains the black hole interior, which in turn contains more and more unpartnered interior Hawking modes as the Hawking radiation escapes into AUX. 

After the Page time, the entanglement wedge of the full boundary slices $\Sigma_{d-1}(t)$ ends at a quantum extremal surface $\gamma_d(t)$ near the horizon~\cite{Pen19}:
\begin{equation}
\mbox{EW}[\Sigma_{d-1}(t)] = \Sigma_d(t) \cap \mbox{Ext}[\gamma_d(t)]~.
\end{equation} 
Here we have chosen $\Sigma_d(t)$ to contain $\gamma_d(t)$, and Ext denotes the spacelike exterior of $\gamma_d$. Since the interior Hawking modes are no longer part of EW$[\Sigma_{d-1}(t)]$, the von Neumann entropy of the entanglement wedge vanishes and so
\begin{equation}
S[\Sigma_{d-1}(t)] = S_{\rm gen}[\mbox{EW}(\Sigma_{d-1}(t))]=\frac{\mathcal{A}[\gamma_d(t)]}{4G_d}~,
\end{equation}
which decreases to zero as the black hole evaporates.

By Eq.~\eqref{complementarity}, EW$[\mbox{AUX}(t)]$ is the complement of EW$[\Sigma_{d-1}(t)]$. Thus, the entropy of AUX will follow the same Page curve. Before the Page time, EW(AUX) only contains AUX, i.e., the early Hawking radiation that has been extracted from the AdS$_d$ spacetime. Its entropy grows as more radiation is produced:
\begin{equation}
   S_{\rm gen}[\mbox{EW(AUX(t))}] = S(\mbox{AUX})~~~~(t<t_{\rm Page})~.   
\end{equation}
After the Page time, EW(AUX) in addition contains an island $I$:
\begin{equation}
    \mbox{EW(AUX(t))} = \mbox{AUX(t)}\cup I(t)~~~,~I= \mbox{Int}(\gamma_d)~,
\end{equation}
where Int denotes the spatial interior of $\gamma_d$ on $\Sigma_d$. $I$ is the black hole interior, which contains Hawking partners that purify the radiation in AUX. Hence, the generalized entropy is then given by the (decreasing) boundary area of this island:
\begin{equation}
    S_{\rm gen}[\mbox{EW(AUX(t))}] = \frac{\mathcal{A}[\gamma_d(t)]}{4G_d}~~~(t>t_{\rm Page})~.
\end{equation}
After the black hole has completely evaporated and all of the Hawking radiation is in AUX, EW(AUX) continues to contain the black hole interior $I$, now a separate ``island universe'' without boundary (see Fig. \ref{fig:3}). 

\subsection{State Paradox and Ensemble Interpretation}
\label{shbap}

In the present setting, the state paradox arises in AUX. On the one hand, the Hawking radiation in AUX is manifestly in a mixed state, whose entropy continues to increase even after the Page time. (In the notation of Ref.~\cite{Almheiri:2019yqk}, this is the ``non-bold state.'') If its entropy did not increase, then there would be no justification for including the black hole interior island in EW(AUX) after the Page time. On the other hand, the generalized entropy of EW(AUX) after the Page time is given by the area of the black hole, which decreases and eventually vanishes. According to the RT prescription, $S_{\rm gen}[\rm{EW}(\rm{AUX})]$ computes the von Neumann entropy of AUX. Hence AUX must be in a different state from what we originally assumed: one whose entropy follows the Page curve. (In the notation of Ref.~\cite{Almheiri:2019yqk}, this is the ``bold state.'') This is a contradiction~\cite{Bousso:2019ykv}.

It is interesting to compare this instantiation of the state paradox to the version that arose in Sec.~\ref{sh}. In Sec.~\ref{sh}, the extrapolate dictionary is used at the last step, to translate the mixed Dyson sphere state to a mixed boundary state, in conflict with the pure state obtained from RT. In the present setting, the extrapolate dictionary is used earlier, when the {\em bulk} Hawking radiation is allowed to escape into AUX by coupling the {\em boundary} to AUX. Strictly it is not possible to couple radiation inside a spacetime to an auxiliary system, since the resulting nonconservation of the stress tensor would violate the Bianchi identity. Thus the coupling is defined through the boundary, and the extrapolate dictionary is used in interpreting this as a transparent boundary condition for the Hawking radiation. As a result of this coupling, the two conflicting quantum states are both in AUX in the end. %So we do not need to appeal to the AdS/CFT dictionary again to establish the state paradox.

As in the previous section, the paradox is resolved if we assume that the bulk calculation computes both the average state (via Hawking's calculation), and the average entropy (via the RT prescription), in an ensemble of unitary boundary theories. The average entropy of $M_{d-1}$ (and also of AUX) follows the Page curve, because it does so in each (unitary) theory. Different members of the ensemble evolve the same initial state to different final states, so the ensemble average of the state is mixed, and {\em its} entropy grows monotonically even after the Page time.  Both sides of the gravity/ensemble duality exhibit a mixed state: in the bulk because we performed Hawking's calculation, and on the boundary because we averaged over the final state produced by different theories. (In the notation of Ref.~\cite{Almheiri:2019yqk}, the ensemble-averaged bold state equals the non-bold state.)

\section{Double Holography Without a Bath}
\label{dh}

Beginning with Ref.~\cite{AMMZ}, a number of interesting papers have explored the RT prescription for evaporating black holes in a ``doubly holographic'' setting~\cite{Rozali:2019day,Chen:2019uhq,Almheiri:2019psy,Sully:2020pza,Bak:2020enw,Chen:2020uac}. The Hawking radiation is mainly carried by excitations of a holographic CFT$_d$ that escape to a (holographic) auxiliary system. The state paradox arises in this setting as well, and we will exhibit it in Sec.~\ref{dhb}. However the analysis is somewhat complicated by the simultaneous appearance of an extra layer of holography and of the auxiliary system. 

In this section, we will separate these two ingredients: we will introduce double holography without an auxiliary system. We will derive an appropriate ``RT-squared'' prescription for computing the von Neumann entropy of the top level CFT$_{d-1}$ from its $d+1$ dimensional doubly holographic bulk dual. We will not analyze black hole evaporation and the state paradox in this section; however, our results will be useful when we do so in Sec.~\ref{dhb}.

\subsection{General Setup}
\label{dh1}

\begin{figure}
    \centering
    \includegraphics[width=0.7\textwidth]{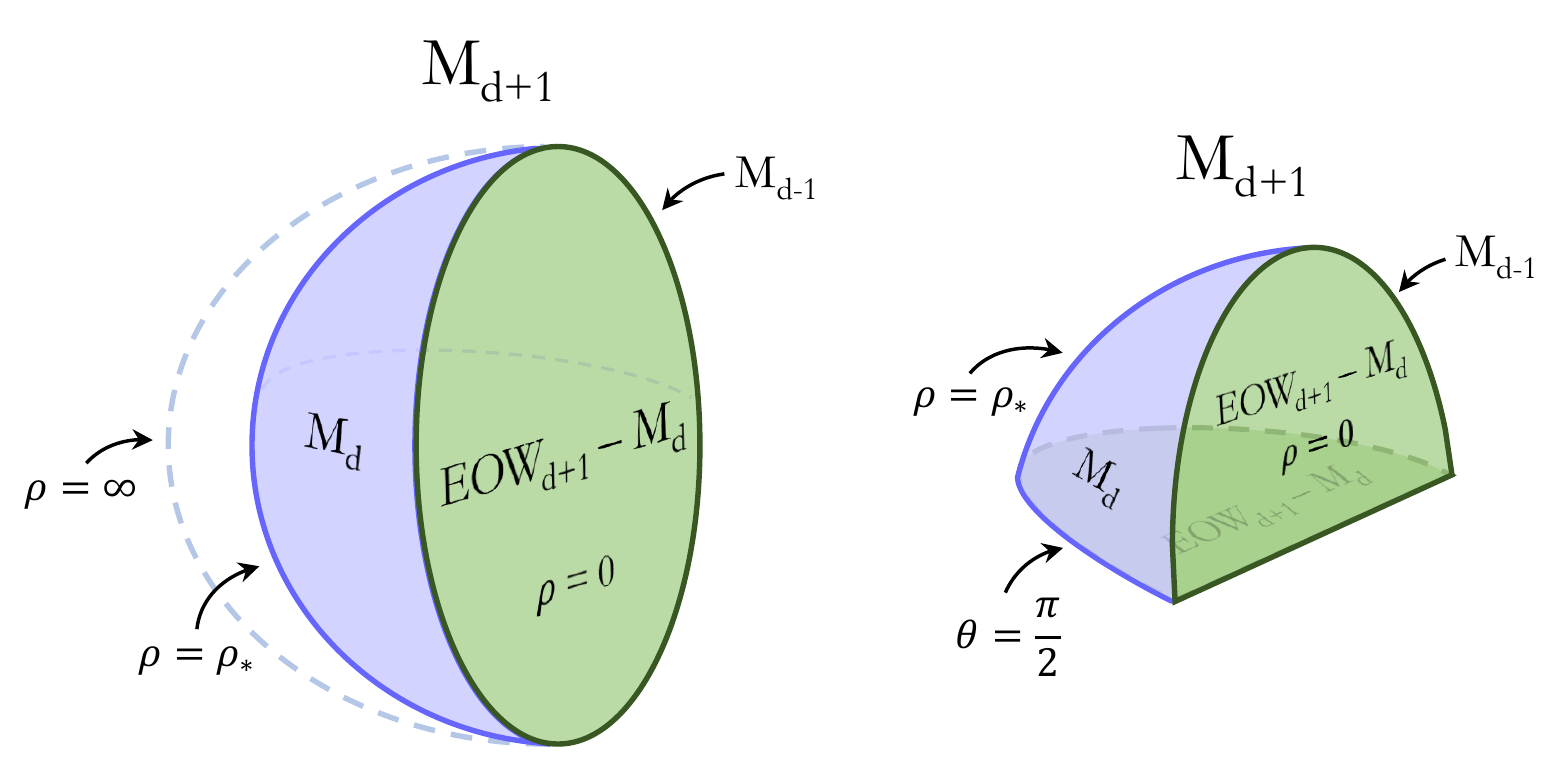}
    \caption{Double holography without a bath. $M_d$ (purple surface) is the bulk dual of a holographic CFT$_{d-1}$ (left) or BCFT$_{d-1}$ (right) on $M_{d-1}$ (dark green rim). So far this is identical to (a time slice of) the setups shown in Fig.~\ref{fig:2.1}. But we now assume that $M_d$ contains a holographic CFT$_d$. This gives rise to a doubly-holographic bulk dual $M_{d+1}$ (the solid interior). From the $d+1$ bulk perspective, $M_d$ is a Karch-Randall braneworld.}
    \label{fig:4.1}
\end{figure}

As in Sec.~\ref{sh}, we consider a holographic CFT$_{d-1}$ on a spacetime $M_{d-1}$, dual to an asymptotically AdS$_d$ spacetime $M_d$:
\begin{equation}
    M_{d-1} \longrightarrow M_d~.
\end{equation}
We now suppose that the matter sector of the $d$-dimensional bulk $M_d$ contains a holographic CFT$_d$ coupled to gravity. This implies that $M_d$ is a Randall-Sundrum braneworld~\cite{RS1,KarRan00}. The holographic duality can then be iterated:
\begin{equation}
    M_d \longrightarrow M_{d+1}~.
\end{equation}
The CFT$_d$ on $M_d$ can be traded for a bulk dual $M_{d+1}$ (see Fig. \ref{fig:4.1}), with Newton's constant $G_{d+1}$ determined by
\begin{equation}
       \frac{G_{d+1}}{L_{d+1}} = \frac{G_d}{d-2}~. \label{gdd}
\end{equation}
Near vacuum regions of the braneworld $M_d$, $M_{d+1}$ will be locally AdS$_{d+1}$, with  curvature length
\begin{equation}
    \frac{L_{d+1}^{d-1}}{G_{d+1}} \sim c_d~.
   \label{gs}
\end{equation}
$M_{d+1}$ will be a manifold with boundary, and we define
\begin{equation}
    \mbox{EOW}_{d+1}=\partial M_{d+1}~.
\end{equation}
By definition, the braneworld $M_d$ is a subset of EOW$_{d+1}$. The complement EOW$_{d+1}-M_d$ is the boundary of the entanglement wedge of the entire AdS$_d$ brane. Therefore it is located at the minimal-area stationary surface anchored on the AdS$_d$ brane's boundary. It implements boundary conditions on the AdS$_{d+1}$ bulk that are dual the reflecting boundary conditions at the boundary of the AdS$_d$ brane. 

The central charge $c_d$ can be thought of as a number of species. In the presence of gravity, large $c_d$ increases the effective Planck length---the cutoff length scale at which the semiclassical analysis breaks down on $M_d$---from $G_d^{1/(d-2)}$ to $(G_d c_d)^{1/(d-2)} \sim L_{d+1}$. 
We assume that $G_{d+1}^{1/(d-1)} \ll L_{d+1}\ll L_d$, or equivalently,
\begin{equation}
    1\ll c_d\ll \frac{L_d^{d-2}}{G_d}~.
\end{equation}
This ensures that $d$-dimensional semiclassical gravity is a valid description both in the $AdS_{d+1}$ bulk (the curvature radius is much greater than the Planck scale) and on the AdS$_d$ brane (the curvature radius is much greater than the cutoff scale $L_{d+1}$).

Usually in holography, there are two descriptions of the same system. The CFT$_{d-1}$ furnishes an exact description. The bulk gives an equivalent description, perturbatively in $G_d$, in the regime where semiclassical gravity (or perturbative string theory) can be applied. In the setting we consider now, there are three levels:
\begin{enumerate}
    \item {\em Top Level:} The CFT$_{d-1}$ on $M_{d-1}$ is the only exact description.
    \item {\em Holographic Bulk Dual:} The asymptotically AdS$_d$ bulk $M_d$ with a CFT$_d$ coupled to gravity is an approximate $d$-dimensional description. Note that this description is {\em alternate} to the CFT$_{d-1}$, so there is no CFT$_{d-1}$ at this level.
    \item {\em Doubly Holographic Bulk Dual:} The third description, also approximate, is $M_{d+1}$. There is no CFT$_d$ on the braneworld, at this level; however any other matter fields and dynamical gravity will still be present on $M_d$.
\end{enumerate}
We will refer to the relation between the top and bottom level as double holography and denote it with a double arrow:
\begin{equation}
    M_{d-1} \Longrightarrow M_{d+1}~.
\end{equation}
Two examples are shown in Fig. \ref{fig:4.1}.

The first example is a holographic CFT$_{d-1}$ on $M_{d-1} = \mathbf{S}^{d-2}\times \mathbf{R}$. In the vacuum state, this is dual to global AdS$_d$. We now take the AdS$_d$ to contain a holographic CFT$_d$ with the above parameters. Then the CFT$_{d-1}$ has a doubly holographic dual which is locally AdS$_{d+1}$:
\begin{gather}
    ds_{d+1}^2 = L_{d+1}^2 \left[ d\rho^2 + \cosh^2\rho\, (-\cosh^2 r dt^2+dr^2+ \sinh^2 r\, d\Omega_{d-2}^2 ) \right]~,
    \label{KRslicing}\\
    0\leq \rho\leq \arccosh \frac{L_d}{L_{d+1}}~.\nonumber
\end{gather}
Here $d\Omega_{d-2}^2=d\theta^2+\sin^2\theta\, d\Omega_{d-3}^2$ is the metric on the unit $d-2$ sphere. In these coordinates, the AdS$_d$ brane $M_d$ sits at $\rho_*$ with $\cosh\rho_*=L_d/L_{d+1}$; a second EOW brane resides at $\rho=0$. See Fig. \ref{fig:4.1}.

The second example of Fig.~\ref{fig:4.1} is half of the previous example. We start with a BCFT$_{d-1}$ on $M_{d-1} = \mathbf{B}^{d-2}\times \mathbf{R}$, where $\mathbf{B}^{d-2}$ is a $d-2$ dimensional hemisphere. For the simplest BCFT with reflecting boundary conditions at the equator, the vacuum state is doubly holographically dual to $M_{d+1}$, the restriction of Eq.~\eqref{KRslicing} to the hemisphere $\theta\leq \pi/2$. There is now an additional EOW$_{d+1}$ at $\theta=\pi/2$. The single holographic dual $M_d$ is half of an AdS$_d$ braneworld (still at $\cosh\rho=L_d/L_{d+1}$), with an EOW$_d$ at $\theta=\pi/2$.

%Note that the CFT$_d$ on the AdS$_d$ brane can be viewed as a ``Boundary" Conformal Field Theory (BCFT)~\cite{Car04,AffLud91}. Because the boundary conditions are simple, the EOW brane has zero tension.

\subsection{One-Step Ryu-Takayanagi Prescription for Double Holography}
\label{dh2}

The von Neumann entropy $S_{d-1}$ of the CFT$_{d-1}$ restricted to an achronal region $R_{d-1}\subset M_{d-1}$ is given by Eq.~\eqref{qhrt}, which we repeat here for convenience:
\begin{equation}
  S(R_{d-1}) = S_{\rm gen}[\mbox{EW}(R_{d-1})]~,
  \label{qhrt2}
\end{equation}
where EW$(R_{d-1})\subset M_d$ is the entanglement wedge. In the doubly holographic setting of this section, $M_d$ is a braneworld.

A Ryu-Takayanagi prescription also applies to braneworlds~\cite{Emp06,MyePou13,KoeLei15}. Let $R_d\subset M_d$ be an achronal region on the braneworld. Then
\begin{equation}
  S_{\rm gen}(R_d) = S_{\rm gen}[\mbox{EW}(R_d)]~.
  \label{rt2}
\end{equation}
More generally, $R_d$ may span both a braneworld region and a region (with no gravity) on $\partial \tilde M_{d+1}$, the conformal boundary of $M_{d+1}$; or it may consist of disconnected components in both types of regions. This case will be important in Sec.~\ref{dhb}; see Fig.~\ref{fig:5.3}, with $R_d=\mbox{EW}(\mathcal{R}_d)$. For $R_d\subset \partial \tilde M_{d+1}$, the generalized entropy on the left hand side is defined as the ordinary von Neumann entropy, with an unregulated UV divergence at $\partial R_d$. Thus Eq.~\eqref{rt2} reduces to the usual RT prescription when $R_d$ is entirely on the true boundary.

The entanglement wedge EW$(R_d)$ is defined as in Sec.~\ref{shrt}, with $d\to d+1$: it is an achronal region $X_{d+1}\subset M_{d+1}$, such that
\begin{enumerate}
\item In the unphysical spacetime, $\partial X_{d+1}=\gamma_{d+1}\cup R_d \cup E_{d+1}$. Here $\gamma_{d+1}\subset M_{d+1}-\mbox{EOW}_{d+1}$, and $E_{d+1}\subset$ EOW$_{d+1}-R_d$. Note that any portion of $R_d$ that lies on a braneworld is a subset of EOW$_{d+1}$.
\item $S_{\rm gen}(X_{d+1})$ is stationary under variations of $\gamma_{d+1}$.
\item $X_{d+1}$ is has the smallest $S_{\rm gen}$ among all regions with the above properties.
\end{enumerate}
Comparing to Eq.~\eqref{qhrt2}, an important modification in Eq.~\eqref{rt2} is that the prescription now computes the {\em generalized} entropy of the region $R_d$, rather than purely a CFT$_d$ von Neumann entropy.

The above rules can be combined iteratively, by choosing $R_d=\mbox{EW}(R_{d-1})$. This allows us to compute any CFT$_{d-1}$ (1st level) entropy using the $d+1$ bulk (the 3rd level). Substituting Eq.~\eqref{rt2} into Eq.~\eqref{qhrt2} we find
\begin{equation}
  S(R_{d-1}) = S_{\rm gen}[\mbox{EW}(\mbox{EW}(R_{d-1}))]~.
  \label{rt12}
\end{equation}
This is a two-step prescription: one first finds the stationary surface $\gamma_d$ on the AdS$_d$ brane, and then one finds the stationary surface $\gamma_{d+1}$ anchored on $\gamma_d$. However, we will now show that this is equivalent to simply minimizing the generalized entropy over surfaces that are allowed to be anchored anywhere on the AdS$_d$ brane (and anywhere on the EOW brane), subject to the homology rules described above. 

To see this, suppose that the latter procedure yielded a surface $\gamma_{d+1}$ whose boundary $\sigma$ on the AdS$_d$ brane was not the minimal QES, $\gamma_d$. Then there are two possibilities: (i) $\sigma$ does not have stationary generalized entropy with respect to small deformations on the brane or (ii) $\sigma$ is stationary but has larger generalized entropy than $\gamma_d$. Case (i) together with the RT rule for braneworlds implies that the generalized entropy of $\gamma_{d+1}$ (in the $d+1$ bulk) is not stationary under small deformations of $\gamma_{d+1}$ that reduce to small deformations of $\sigma$. Case (ii) implies that the $d+1$ bulk stationary surface anchored on $\gamma_d$ has smaller generalized entropy than $\gamma_{d+1}$. Either of these implications contradicts the definition of $\gamma_{d+1}$.

Thus we can formulate a {\bf one-step Ryu-Takayanagi prescription} for the von Neumann entropy of a region $R_{d-1}$ of a doubly-holographic CFT$_{d-1}$:
\begin{equation}
    S(R_{d-1})= S_{\rm gen}[\mbox{EW}^2(R_{d-1})]~,
    \label{onestepdh}
\end{equation}
where EW$^2(R_{d-1})$ denotes the {\em doubly-holographic entanglement wedge} of $R_{d-1}$. This is defined as an achronal region $X_{d+1}\subset M_{d+1}$ such that 
\begin{enumerate}
\item In the unphysical spacetime, $\partial X_{d+1}=R_{d-1}\cup \gamma_{d+1}\cup E_{d+1}$. Here $\gamma_{d+1}\subset M_{d+1}-\mbox{EOW}_{d+1}$ and $E_{d+1}\subset$ EOW$_{d+1}$.
\item $S_{\rm gen}(X_{d+1})$ is stationary under variations of $\gamma_{d+1}$.
\item $X_{d+1}$ has the smallest $S_{\rm gen}$ among all regions with the above properties.
\end{enumerate}

\begin{figure}
    \centering
    \includegraphics[width=\textwidth]{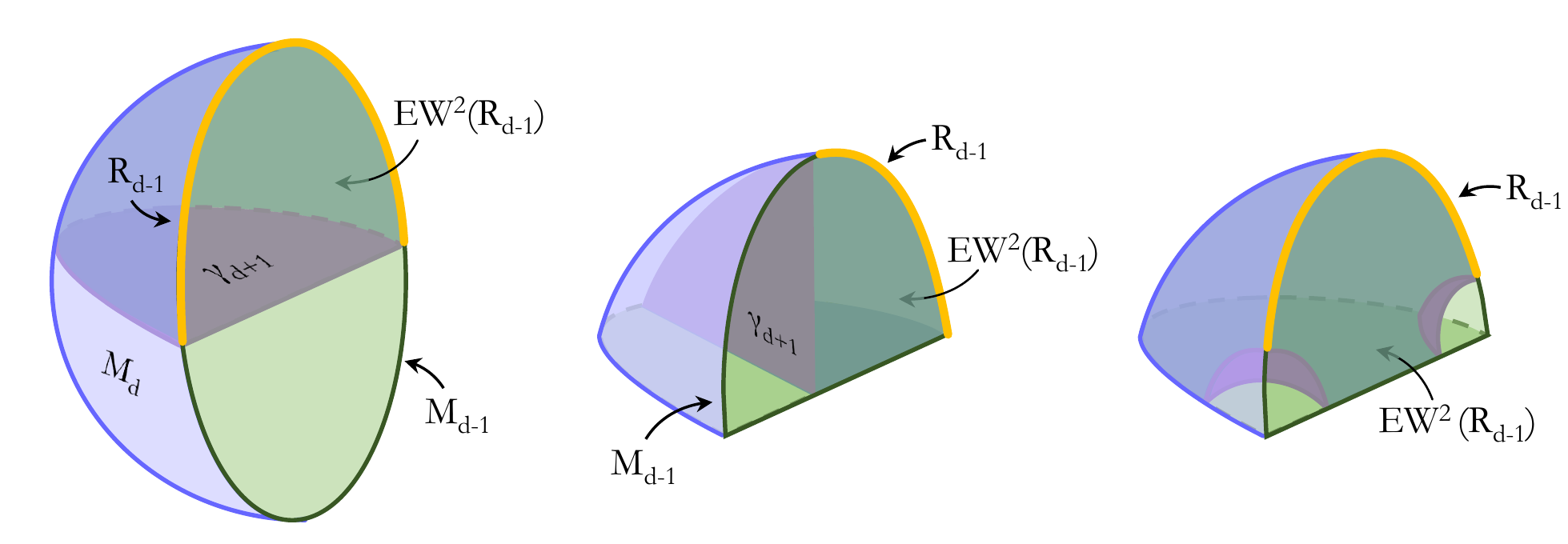}
    \caption{Examples of the doubly holographic entanglement wedge $\mbox{EW}^2(R_{d-1})$ for a (B)CFT$_{d-1}$ region $R_{d-1}$. As before, the light purple surface $M_d$ is the bulk dual of a holographic CFT$_{d-1}$ (left) or BCFT$_{d-1}$ (middle, right) on $M_{d-1}$ (dark green rim).  In each case, the doubly holographic entanglement wedge is bounded in part by the surface $\gamma_{d+1}$, shown in dark purple.}
    \label{fig:4.2}
\end{figure}

A very simple example is shown in Fig.~\ref{fig:4.2}. Consider the CFT$_{d-1}$ in the vacuum state, and let $R$ be half of the $d-1$ sphere in standard global coordinates. Then the QES $\gamma_d$ is a $d-1$ dimensional hyperbolic plane cutting a Cauchy surface of the AdS$_d$ brane in half: $\cosh \rho = L_d/L_{d+1}\,;~\theta=\pi/2$. (In this example the quantum corrections play no role, so this is also a classical stationary surface.) The QES $\gamma_{d+1}$ is similarly part of a hyperbolic plane cutting the Cauchy surface of the AdS$_{d+1}$ bulk in half: $\theta=\pi/2$. Of course, it only includes the portion between the AdS$_d$ brane and the EOW brane: $1<\cosh\rho<L_d/L_{d+1}$. Fig.~\ref{fig:4.2} also shows other examples.

\subsection{Quantum vs.\ Classical RT in Double Holography}
\label{qc}

In the case where the generalized entropies of $\gamma_d$ and $\gamma_{d+1}$ are both dominated by the area terms, consistency of Eqs.~\eqref{rt2} and \eqref{rt12} requires 
\begin{equation}
    \frac{{\cal A}(\gamma_d)}{4G_d} = \frac{{\cal A}(\gamma_{d+1})}{4G_{d+1}}~;
\end{equation}
By Eq.~\eqref{gdd}, this implies a very simple relation between the areas of the QESs:
\begin{equation}
    {\cal A}(\gamma_{d+1}) = {\cal A}(\gamma_d) \, \frac{2L_{d+1}}{d-2}~.
\end{equation}
It is easy to check that this relation is obeyed in the above examples. More generally, consistency requires that $\gamma_{d+1}$ must have a phase transition if and only if $\gamma_d$ does, as the region $R$ is varied. For example, if $R$ consists of two antipodal round disks of equal size in the CFT$_{d-1}$, then $\gamma_d$ undergoes a well-known phase transition as the disk radius is varied. $\gamma_{d+1}$ must also have a phase transition at the same critical radius. At first this behavior may seem surprising, because one expects the QESs in the $d+1$ bulk to have a richer structure than those on the AdS$_d$ brane. However, in this context we are only considering $d+1$ QESs anchored on very special surfaces on the AdS$_d$ brane---those that are themselves QESs---so there is no contradiction.

A more interesting case arises when the CFT$_d$ is far from its vacuum state, so that the von Neumann entropy of braneworld regions is large. In this case $S_{\rm gen}$ on $M_d$ may have large quantum contributions (i.e., contributions from the von Neumann entropy term), while $S_{\rm gen}$ of the corresponding entanglement wedge in $M_{d+1}$ is dominated by the classical term (the area term). In such a case, one can replace $S_{\rm gen}$ by $\mathcal{A}(\gamma_{d+1})/4G_{d+1}$ in Eqs.~\eqref{rt2}--\eqref{onestepdh}, but not by $\mathcal{A}(\gamma_d)/4G_d$ in Eq.~\eqref{qhrt2}.  

\section{Double Holography With a Holographic Bath}
\label{dhb}

This section can be thought of as an extension of the above settings, in two different ways. Continuing from the previous section, we keep the doubly holographic setup but we add a bath. That is, we couple the CFT$_{d-1}$ (or equivalently, the AdS$_d$ brane) to an auxiliary system AUX. We take AUX to be the same holographic CFT$_d$ that lives on the AdS$_d$ brane, but not coupled to gravity. Thus AUX can be thought of as a CFT$_d$ living on a true asymptotic boundary of an asymptotically AdS$_{d+1}$ bulk dual. 

From the perspective of Sec.~\ref{shb}, we keep the bath but make the setting doubly holographic. That is, we now specialize to the case where both the dominant matter content in the gravitating AdS$_d$ spacetime, and also the external bath AUX is a holographic CFT$_d$, with an asymptotically AdS$_{d+1}$ bulk dual.

Combining insights from the previous sections will allow us to understand some puzzling features in the doubly-holographic versions~\cite{AMMZ,Rozali:2019day,Chen:2019uhq,Almheiri:2019psy,Sully:2020pza,Bak:2020enw} of Refs.~\cite{Pen19,AEMM}, where the Page curve arises from the classical RT prescription in the $d+1$ dimensional bulk. One such feature is the appearance of two apparently different states in the bath region, denoted bold and non-bold in Ref.~\cite{AMMZ}. We will see that these states need not be different in the ensemble interpretation.

\begin{figure}
    \centering
    \includegraphics[width=.6\textwidth]{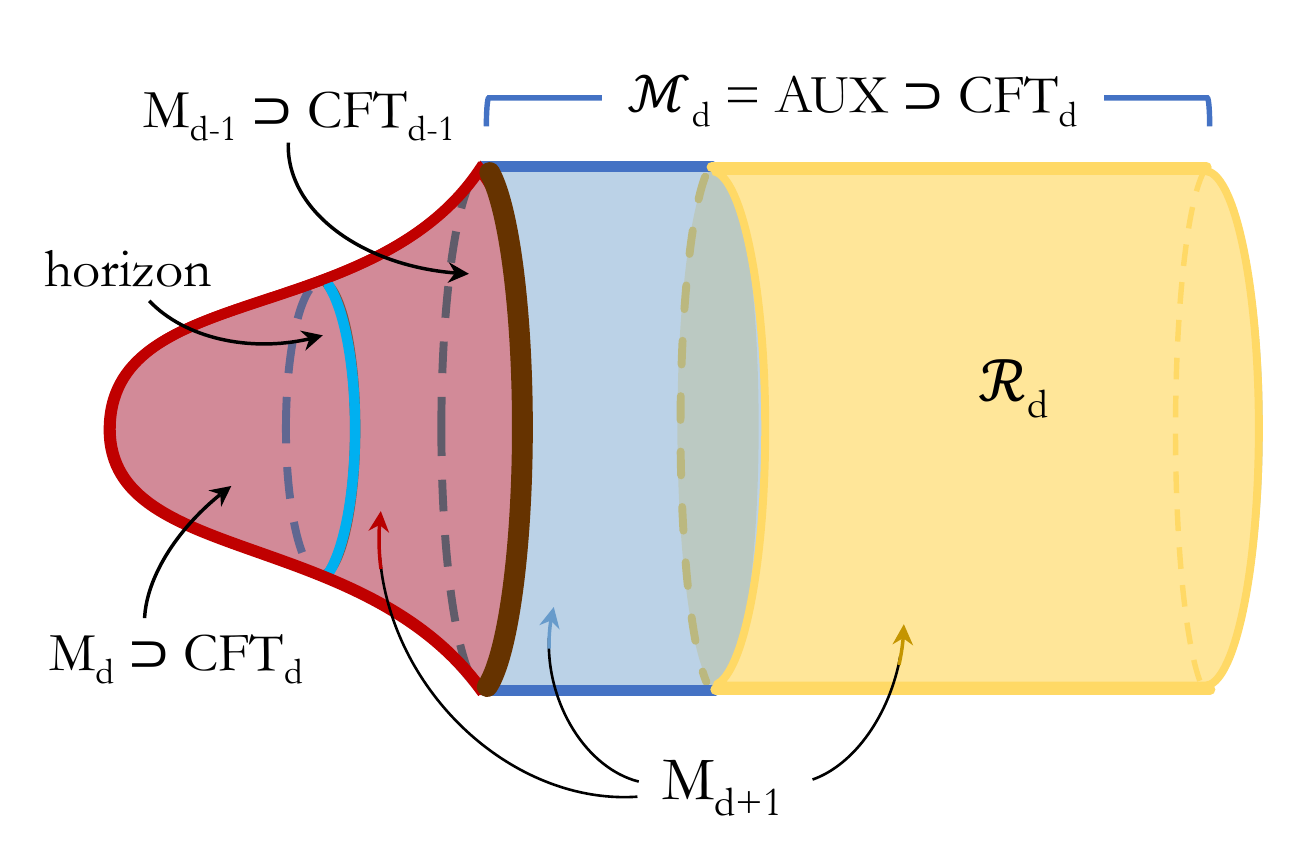}
    \caption{A doubly holographic CFT$_{d-1}$ on $M_{d-1}$ is coupled to holographic bath: a CFT$_d$ on $\mathcal{M}_d$. The first holographic dual is $M_d\cup \mathcal{M}_d$, where $M_d$ contains the same CFT$_d$ coupled to gravity. The second holographic dual is $M_{d+1}$ (solid interior). We consider a state which, in the first dual, corresponds to an evaporating black hole in $M_d$ with radiation escaping to $\mathcal{M}_d$. The von Neumann entropy of the radiation in the subregion $\mathcal{R}_d\subset \mathcal{M}_d$ can be computed using the single or double RT prescription.}
    \label{fig:5.1}
\end{figure}

\subsection{General Setup}
\label{dhb1}

As before, we consider a holographic CFT$_{d-1}$ with central charge $c_{d-1}$ on a manifold $M_{d-1}$, dual to a bulk $M_d$. We now choose this CFT$_{d-1}$ such that the matter content of $M_d$ includes a particular CFT$_d$ coupled to gravity. As in Sec.~\ref{shb}, we couple the CFT$_{d-1}$ to an auxiliary system AUX (see Fig. \ref{fig:5.1}). We now insist that AUX is specifically a CFT$_d$ on a manifold $\mathcal{M}_d$ such that $M_{d-1}=\partial \mathcal{M}_d$, and we take this to be the same CFT$_d$ that also appears in the bulk dual $M_d$. 

The coupled boundary system (CFT$_{d-1}$ on $M_{d-1}$ and CFT$_d$ on $\mathcal{M}_d$) defines a BCFT$_d$ on $\mathcal{M}_d$. Importantly, there is no dynamical gravity on $\mathcal{M}_d$. Applying the general discussion of Sec.~\ref{shb} to the CFT$_{d-1}$ and AUX (i.e., to the BCFT$_d$), we find that this system is holographically dual to a $d$-dimensional bulk system: 
\begin{equation}
     M_{d-1}\cup \mathcal{M}_d ~~~\longrightarrow~~~M_d\cup \mathcal{M}_d~.
    \label{firstdb}
\end{equation} 
Here $M_d$ has dynamical gravity. $\mbox{AUX}=\mathcal{M}_d$ plays a dual role as bulk and boundary system. 

Next, we add the ingredient of double holography, as in Sec.~\ref{dh}. Suppose that the CFT$_d$ on $M_d\cup \mathcal{M}_d$ is holographic, with parameters as described in Sec.~\ref{dh1}. Let $M_{d+1}$ be its $d+1$ dimensional bulk dual: 
\begin{equation}
    M_d\cup \mathcal{M}_d~~~\longrightarrow~~~M_{d+1} ~.
    \label{seconddb}
\end{equation} 
As usual, let $\tilde M_{d+1}$ be the associated unphysical spacetime (Penrose diagram), and let EOW$_{d+1}=\partial M_{d+1}$. Then $\mathcal{M}_d=\partial \tilde M_{d+1}$ and $M_d\subset$ EOW$_{d+1}$. The above two dualities combine to establish the doubly holographic duality
\begin{equation}
    M_{d-1}\cup \mathcal{M}_d~~~\Longrightarrow~~~M_{d+1} ~.
    \label{doubledb}
\end{equation} 

For example, with $M_{d-1}=\mathbf{S}^{d-2}\times \mathbf{R}$ at the equator of the hemishere $\mathcal{M}_d=\mathbf{B}^{d-1}\times \mathbf{R}$, one obtains the Karch-Randall (KR) model~\cite{KarRan00}. This was first discussed in detail as a doubly-holographic model in Ref.~\cite{BouRan01}. The first bulk dual is $M_d\cup \mathcal{M}_d$, where $M_d$ is an AdS$_d$ braneworld known as a KR brane. It forms the boundary of the doubly holographic dual $M_{d+1}$, a global AdS$_{d+1}$ spacetime that terminates on the KR brane. In the vacuum state, the metric of $M_{d+1}$ is given by Eq.~\eqref{KRslicing}, with the range of $\rho$ extended to
\begin{equation}
        -\infty < \rho\leq \arccosh \frac{L_d}{L_{d+1}}~;
\end{equation}
The braneworld $M_d$ is located at the upper end of this range, and the asymptotic boundary $\mathcal{M}_d$ is at the lower end. $M_{d-1}$ is at $\rho=0$, $r\to\infty$.

Alternatively, let $M_{d-1}=\mathbf{R}^{d-2}\times \mathbf{R}$ be the boundary of the half-space $\mathcal{M}_d=\mathbf{B}^{d-1}\times \mathbf{R}$. This gives the Poincare patch of an AdS$_d$ braneworld as the first bulk dual, $M_d$; it gives the Poincare patch of AdS$_{d+1}$ as the second bulk dual $M_{d+1}$. 

Both of these models were studied further by Takayanagi and collaborators~\cite{Tak11,FujTak11}, who gave a one-step RT prescription for the duality in Eq.~\eqref{doubledb}. We will now derive this prescription from a different perspective, by combining the results of the previous sections. 

\subsection{One-Step Ryu-Takayanagi prescription for Double Holography}
\label{dhb2}

The one-step RT prescription for the doubly holographic duality \eqref{doubledb} can be derived iteratively by combining the RT prescriptions for the single holographic dualities \eqref{firstdb}  and \eqref{seconddb}. For the first step this was given in Eqs.~\eqref{island}-\eqref{ewshb}. Setting AUX $\to \mathcal{R}_d$, Eq.~\eqref{island} becomes:
\begin{equation}
  S(R_{d-1}\cup \mathcal{R}_d) = S_{\rm gen}[\mbox{EW}(R_{d-1}\cup \mathcal{R}_d)]~,
  \label{island2}
\end{equation}
where $R_{d-1}\subset M_{d-1}$ and $\mathcal{R}_d\subset \mathcal{M}_d-M_{d-1}$ are arbitrary subregions of the boundary system. The other equations and the definition of EW are as in Sec.~\ref{shb}. The fact that the auxiliary system is a field theory plays no role in this step.

The second step computes the generalized entropy on the RHS of Eq.~\eqref{island2} holographically. Setting $R_d \to \mbox{EW}(R_{d-1}\cup \mathcal{R}_d)$ in Eq.~\eqref{rt2}, we obtain
\begin{equation}
  S_{\rm gen}(\mbox{EW}(R_{d-1}\cup \mathcal{R}_d)) = S_{\rm gen}[\mbox{EW}(\mbox{EW}(R_{d-1}\cup \mathcal{R}_d))]~.
  \label{rt22}
\end{equation}
Thus we obtain
\begin{equation}
  S(R_{d-1}\cup \mathcal{R}_d) = S_{\rm gen}[\mbox{EW}(\mbox{EW}(R_{d-1}\cup \mathcal{R}_d))]~.
 \label{rt3}
\end{equation}

By arguments exactly analogous to those following Eq.~\eqref{rt12}, this iterative result can be condensed into a {\bf one-step RT prescription}:
%Let $M_{d-1}=\partial \mathcal{M}_d$; and let BCFT$_d$ be a Boundary CFT on $\mathcal{M}_d$, obtained by coupling a doubly holographic CFT$_{d-1}$ on $M_{d-1}$ to a holographic CFT$_d$ on $\mathcal{M}_d-M_{d-1}$. For simplicity we assume that the first holographic dual to the CFT$_{d-1}$, Eq.~\eqref{firstdb} is a spacetime $M_d$ whose matter includes the same CFT$_d$, coupled to gravity. Then the doubly holographic duality of Eq.~\eqref{doubledb} can be used to compute the von Neumann entropy of an arbitrary BCFT$_d$ subregion, consisting of $R_{d-1}\subset M_{d-1}$ and $\mathcal{R}_d\subset \mathcal{M}_d-M_{d-1}$:
\begin{equation}
  S(R_{d-1}\cup \mathcal{R}_d) = S_{\rm gen}[\mbox{EW}^2(R_{d-1}\cup \mathcal{R}_d)]~.
 \label{rt4}
\end{equation}
The doubly-holographic entanglement wedge EW$^2(R_{d-1}\cup \mathcal{R}_d)$ is defined as an achronal region $X_{d+1}\subset M_{d+1}$ such that 
\begin{enumerate}
\item In the unphysical spacetime, $\partial X_{d+1}=R_{d-1}\cup \mathcal{R}_d \cup \gamma_{d+1}\cup E_{d+1}$, where $\gamma_{d+1}\subset M_{d+1}-\mbox{EOW}_{d+1}$ and $E_{d+1}\subset$ EOW$_{d+1}$.
\item $S_{\rm gen}(X_{d+1})$ is stationary under variations of $\gamma_{d+1}$.
\item $X_{d+1}$ is has the smallest $S_{\rm gen}$ among all regions with the above properties.
\end{enumerate}
We note that this agrees with the RT prescription for a BCFT$_d$ given by Takayanagi~\cite{Tak11,FujTak11}, which has been extensively used in recent analyses of entanglement islands, such as Refs.~\cite{AMMZ,Rozali:2019day,Chen:2019uhq,Almheiri:2019psy,Sully:2020pza,Bak:2020enw}. In analyzing these results and exhibiting the state paradox, it will be illuminating to ``deconstruct'' Eq.~\eqref{rt4} in to Eqs.~\eqref{island2} and \eqref{rt22}.

\subsection{Island and Page Curve}
\label{dhbip}

\begin{figure}
    \centering
    \includegraphics[width=\textwidth]{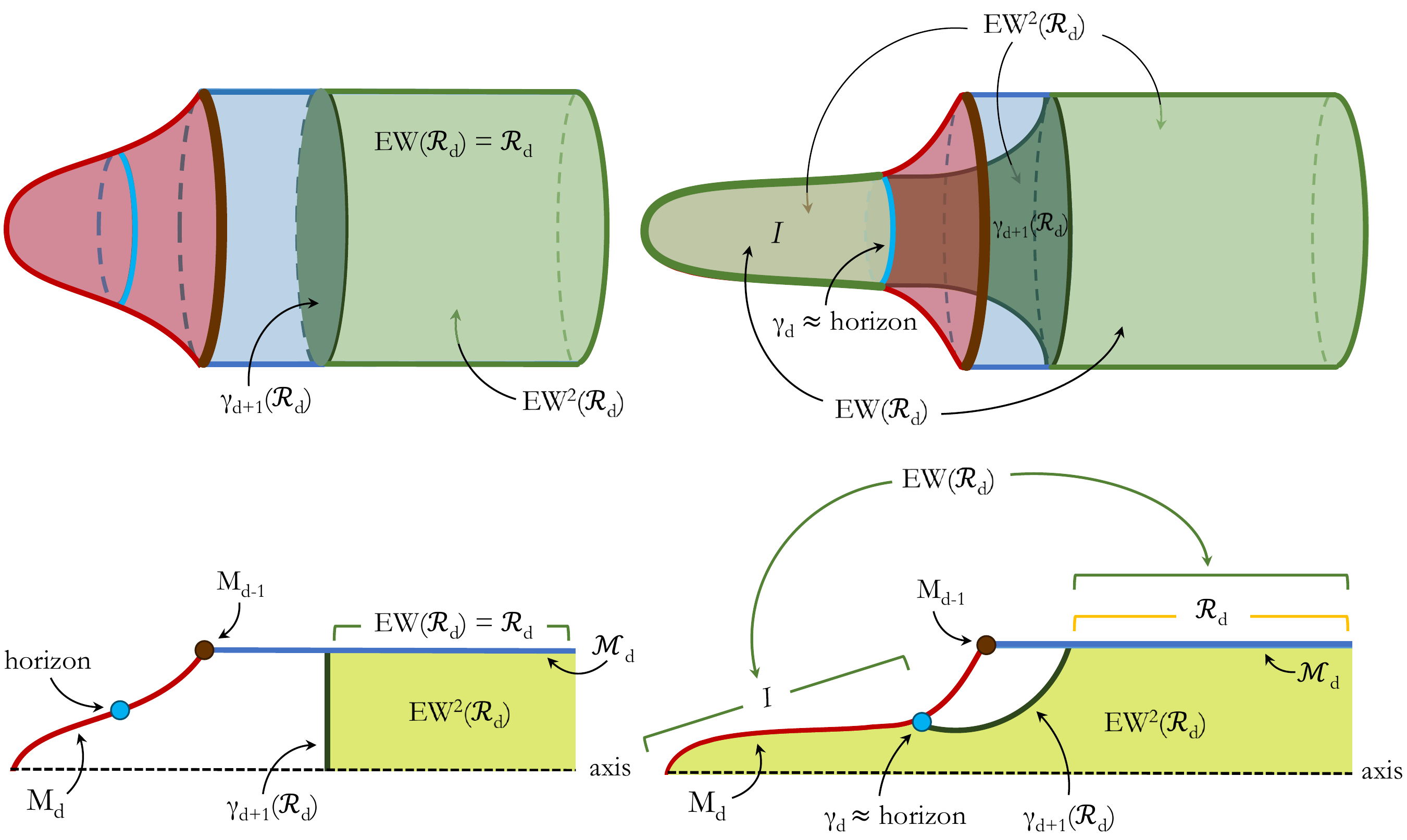}
    \caption{Entanglement wedges of the bath region $\mathcal{R}_d$ before (left) and after (right) the Page time, when EW$(\mathcal{R}_d)$ has a disconnected island $I$. Each top figure is simply the bottom figure rotated around the axis. The ``squared'' entanglement wedge EW$^2$ is always connected. It can be found iteratively as EW(EW$(\mathcal{R}_d)$), or in one step from Eq.~\eqref{rt4}~\cite{Tak11,FujTak11}. As in Sec.~\ref{dhbip}, $\gamma_d$ is a quantum extremal surface, but $\gamma_{d+1}$ is an ordinary extremal surface.}
    \label{fig:5.3}
\end{figure}

We now specialize to the dynamical setting of Sec.~\ref{shbip}: the first holographic dual, $M_d\cup \mathcal{M}_d$, $M_d$ contains a black hole whose radiation propagates to $\mathcal{M}_d$. First, let us consider the top level, the BCFT$_d$ on $M_{d-1}\cup\mathcal{M}_d$. There are now two ways to compute the von Neumann entropy of a subregion $\mathcal{R}_d\subset\mathcal{M}_d$ that contains the radiation. 

One option is to ignore the second holographic dual and use RT only for the first holographic duality, Eq.~\eqref{firstdb}. Setting $R_{d-1}\to\varnothing$ 
in Eq.~\eqref{island2}, we find
\begin{equation}
    S(\mathcal{R}_d) = S_{\rm gen}[\mbox{EW}(\mathcal{R}_d)]~.
    \label{srd1}
\end{equation}
Before the Page time, $\mbox{EW}(\mathcal{R}_d)=\mathcal{R}_d$ (see Fig.~\ref{fig:5.3}). Since $\mathcal{R}_d$ is a true boundary region, $S_{\rm gen}(\mathcal{R}_d) = S(\mathcal{R}_d)$. Thus, the above equation is a trivial identity before the Page time. After the Page time,
\begin{equation}
   \mbox{EW}(\mathcal{R}_d) = \mathcal{R}_d\cup I ~~~~(t>t_{\rm Page})~,
   \label{ew1}
\end{equation}
where the island $I\subset M_d$ is the black hole interior (see Fig.~\ref{fig:5.3}). The Hawking radiation in $\mathcal{R}_d$ is purified by the Hawking partners in $I$, so 
\begin{equation}
S_{\rm gen}[\mbox{EW}(\mathcal{R}_d)]= \frac{\mathcal{A}(\gamma_d)}{4G_d} ~~~~(t>t_{\rm Page})~,    
\end{equation} 
where $\gamma_d=\partial I$ nearly coincides with the horizon. Note that the radiation appears on both sides of the duality, and that we have made no reference to the second holographic bulk dual $M_{d+1}$.

Another option is to use the doubly holographic duality, Eq.~\eqref{doubledb}. By Eq.~\eqref{rt4},
\begin{equation}
    S(\mathcal{R}_d) = S_{\rm gen}[\mbox{EW}^2(\mathcal{R}_d)]~.
    \label{srd2}
\end{equation}
With the one-step prescription following Eq.~\eqref{rt4} one finds EW$^2(\mathcal{R}_d)$ as shown in Fig.~\ref{fig:5.3}. Unlike EW$(\mathcal{R}_d)$ in Eq.~\eqref{ew1}, $\mbox{EW}^2(\mathcal{R}_d)$ is always a connected region. After the Page time, $\gamma_{d+1}$ ends on the quantum extremal surface $\gamma_d$, and the island $I$ forms part of the boundary of $\mbox{EW}^2(\mathcal{R}_d)$. But neither the radiation in $\mathcal{R}_d$ nor the Hawking partners in the black hole interior on $M_d$ contribute to $S_{\rm gen}[\mbox{EW}^2(\mathcal{R}_d)]$, since they are not part of $M_{d+1}$. Both before and after the Page time, the generalized entropy of the squared entanglement wedge is given just by the classical area of $\gamma_{d+1}$, in line with the discussion at the end of Sec.~\ref{qc}:
\begin{equation}
     S_{\rm gen}[\mbox{EW}^2(\mathcal{R}_d(t))]=\frac{\mathcal{A}[\gamma_{d+1}(t)]}{4G_{d+1}}~.
     \label{areaonly}
\end{equation}

\subsection{State Paradox and Ensemble Interpretation}
\label{dhbap}

Agreement between Eqs.~\eqref{srd1} and \eqref{srd2} is a nontrivial consequence of Eq.~\eqref{rt22}. That equation, in turn, was obtained by applying the RT prescription for braneworlds, \eqref{rt2}, which is relevant for the duality~\eqref{seconddb}, to the region EW$(\mathcal{R}_d)$. But Eq.~\eqref{rt2} allows us to choose any other subregion of the first bulk dual $M_d\cup\mathcal{M}_d$ and compute its generalized entropy. Thus we may ask questions that have no obvious analogue in the dualities of Eqs.~\eqref{firstdb} and \eqref{doubledb}. 

\begin{figure}
    \centering
    \includegraphics[width=\textwidth]{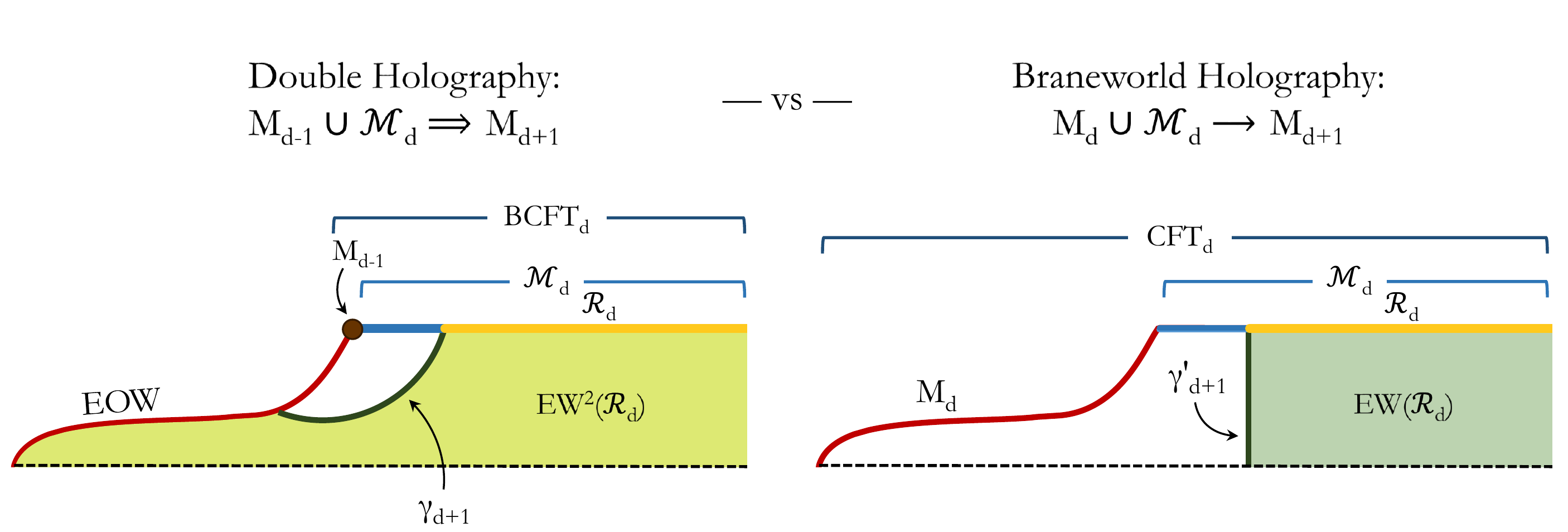}
    \caption{Bulk dual of the state paradox. Left: we regard $\mathcal{R}_d$ as a BCFT$_d$ subregion (top level). The homology rule following Eq.~\eqref{rt4} applies: $\gamma_{d+1}$ is allowed to end on the braneworld which here appears as an EOW brane. At late times, $\mathcal{A}(\gamma_{d+1})\to 0$, resulting in the Page curve for $S(\mathcal{R}_d)$. Right: we consider $\mathcal{R}_d$ as a subregion of the CFT$_d$ on $M_d\cup \mathcal{M}_d$. The homology rule following Eq.~\eqref{rt2} applies. The braneworld $M_d$ is now part of the boundary; since we are computing the entropy only for the region $\mathcal{R}_d$, $\gamma'_{d+1}$ is not allowed to end on $M_d$. $\mathcal{A}(\gamma'_{d+1})$ grows monotonically, resulting in Hawking's curve.}
    \label{fig:5.4}
\end{figure}

For example, after the Page time, EW$(\mathcal{R}_d)= \mathcal{R}_d\cup I$. But we could instead use Eq.~\eqref{rt2} to compute the generalized entropy of just $\mathcal{R}_d$. Because Eq.~\ref{rt2} prohibits the RT surface $\gamma_{d+1}'$ from ending on $M_d$ (see Fig.~\ref{fig:5.4}), its area continues to grow after the Page time, and we find the entropy computed by Hawking. Thus, Eq.~\eqref{rt2} will not give the same answer for $S(\mathcal{R}_d)$ as Eqs.~\eqref{srd1} and \eqref{srd2}! This contradiction is the bulk dual of the state paradox.

In Sec.~\ref{shbap} (with AUX $\to \mathcal{R}_d$), the state paradox appeared as a contradiction between $S(\mathcal{R}_d)$ computed from the semiclassical Hawking analysis on $M_d\cup\mathcal{M}_d$, and $S(\mathcal{R}_d)$ computed from Eq.~\eqref{srd1}. Either quantity can now also be computed using the second holographic dual $M_{d+1}$. As noted in the previous paragraph, the results (given by Eq.~\eqref{rt2} and~\eqref{srd2} respectively) disagree.

Gravity/ensemble duality can again resolve this paradox. Suppose that the CFT$_{d-1}$ on $M_{d-1}$ is really an ensemble of unitary theories as discussed in the introduction. From the top-level viewpoint, the CFT$_{d-1}$ emits radiation into the CFT$_d$ on $\mathcal{M}_d$. In each theory, this process is unitary and the radiation entropy in $\mathcal{R}_d\subset\mathcal{M}_d$ follows the Page curve. Hence the average entropy $\braket{S(\rho_{\mathcal{R}_d})}$ follows the Page curve. But the ensemble-averaged state of the radiation, $S(\braket{\rho_{\mathcal{R}_d}})$, follows Hawking's monotonically rising curve.

The first holographic dual of this process is the escape of Hawking radiation from $M_d$ into $\mathcal{M}_d$. Assuming gravity/ensemble duality, the semiclassical analysis of black hole evaporation computes $\braket{\rho_{\mathcal{R}_d}}$ directly, and it determines $\braket{S(\rho_{\mathcal{R}_d})}$ via the first RT prescription, Eq.~\eqref{srd1}. The second layer of holography, Eq.~\eqref{seconddb}, gives us an alternative way of computing $\braket{S(\rho_{\mathcal{R}_d})}$ and $S(\braket{\rho_{\mathcal{R}_d}})$ using the braneworld version of the RT prescription, Eq.~\eqref{rt2}. To compute $\braket{S(\rho_{\mathcal{R}_d})}$, choose $R_d\to \mbox{EW}(\mathcal{R}_d)=\mathcal{R}_d\cup I$ in Eq.~\eqref{rt2}. To compute $S(\braket{\rho_{\mathcal{R}_d}})$, set $R_d\to \mathcal{R}_d$ in Eq.~\eqref{rt2}. 

It is interesting to note that it does not matter whether Eq.~\eqref{seconddb} is a gravity/ensemble duality. Suppose that it is. Then there exists an ensemble of CFT$_d$ theories on $M_d\cup \mathcal{M}_d$. On what is now the boundary side, we would have to perform a gravity path integral involving each of these different theories, then average. But regardless of the details of each CFT$_d$, the state in $\mathcal{R}_d$ will be thermal and purified by the excitation in $I$. Therefore, unlike the state of the Hawking radiation in $\mathcal{R}_d$ in the BCFT$_d$ (the top level), the state of the semiclassically evolved CFT$_d$ theories is self-averaging in the region $\mathcal{R}_d\cup I$, and also in the region $\mathcal{R}_d$. Of course, in a different state (for example, a setup analogous to Sec.~\ref{sh} in the $d+1$ dimensional bulk), a state paradox can arise in $M_d\cup \mathcal{M}_d$, and we would need to appeal to state gravity/ensemble duality for a resolution.

So far, we have discussed the first and second holographic duality separately. We can also consider the one-step doubly holographic RT prescription of Eq.~\eqref{rt4}. This evaluates the entropy of the BCFT$_d$ region $\mathcal{R}_d$ directly in the $M_{d+1}$ bulk as the area of $\gamma_{d+1}$; see Eq.~\eqref{areaonly}. By ``jumping'' over the middle level, we have missed the paradox. Namely, the paradox involved the apparent discrepancy of the states in the region $\mathcal{R}_d$, depending on whether it is viewed as a state of the BCFT$_d$ or a state of the CFT$_d$ on $M_d\cup \mathcal{M}_d$. The CFT$_d$ is not present in the second holographic dual. It has now been replaced by the classical bulk state in $M_{d+1}$; thus we are no longer comparing two states of the same region.

Therefore, an ensemble interpretation is not required to make sense of the doubly holographic duality \eqref{doubledb}, so long as we never consider the intermediate level. Unfortunately, without the intermediate level $M_d\cup \mathcal{M}_d$, we also lose contact with the process of black hole evaporation, which is manifest only at this level. 

\noindent {\bf Acknowledgements.}~We would like to thank C.~Akers, A.~Almheiri, N.~Bao, V.~Chan\-dra\-sekaran, R.~Emparan, N.~Engelhardt, T.~Hartman, A.~Levine, R.~Mahajan, J.~Maldacena, A.~Maloney, H.~Marrochio, H.~Maxfield, M.~Miyaji, G.~Penington, A.~Shah\-bazi-Moghaddam, R.~Myers, P.~Rath, G.~Remmen, J.~Santos, D.~Stanford, M.~Toma\v{s}evi\'{c} and Y.~Zhao for helpful discussions, comments, and correspondence.  This work was supported in part by the Berkeley Center for Theoretical Physics; by the Department of Energy, Office of Science, Office of High Energy Physics under QuantISED Award DE-SC0019380 and under contract DE-AC02-05CH11231; and by the National Science Foundation under grant PHY1820912.

\bibliographystyle{utcaps}
\bibliography{DH}
\end{document}